\documentclass[fleqn,usenatbib]{mnras}

\usepackage{newtxtext,newtxmath}
\usepackage[T1]{fontenc}

\DeclareRobustCommand{\VAN}[3]{#2}
\let\VANthebibliography\thebibliography
\def\thebibliography{\DeclareRobustCommand{\VAN}[3]{##3}\VANthebibliography}

\usepackage{graphicx} 
\usepackage{amsmath}	% Advanced maths commands
\usepackage[super]{nth} % Correct labeling of occurances, e.g. 1^st 3^rd
\usepackage{siunitx}    % For units
\usepackage{bm}         % For bold math
\usepackage{datatool,booktabs,pgfplotstable} %Table Formating, First two are a requirement of the last
\pgfplotsset{compat=1.18} %!warning if not
\pgfplotstableset{
    every head row/.style={before row=\toprule, after row=\midrule},
    every last row/.style={after row=\bottomrule}
}%top, mid and bottom rule by default
\DeclareSIUnit\angstrom{\text{Å}} % Angstrom 
\DeclareSIUnit\year{yr} % Year
\DeclareSIUnit\Msun{{\rm M}_{\sun}} % Solar Mass
\DeclareSIUnit\pc{pc} % Parsec
\makeatletter
\let\oldtheequation\theequation
\renewcommand\tagform@[1]{\maketag@@@{\ignorespaces#1\unskip\@@italiccorr}}
 
\renewcommand\theequation{(\oldtheequation)}
\makeatother
\newcommand{\GG}[1]{}%To get ordering right of bib

\title[Origin of high {[Ba/Fe]} in NGC1569-B]{Chemical enrichment by collapsars as the origin of the unusually high [Ba/Fe] in a massive star cluster of the dwarf galaxy NGC 1569}

\author[B. Leicester et al.]{
Brayden Leicester,$^{1,2}$\thanks{E-mail: brayden.leicester@canterbury.ac.nz (BL)} 
Kenji Bekki$^{1}$
Takuji Tsujimoto$^{3}$
\\
$^{1}$ICRAR M468 The University of Western Australia 35 Stirling Hwy, Crawley Western Australia 6009, Australia\\
$^{2}$School of Physical and Chemical Sciences  --- Te Kura Mat\={u}, University of Canterbury, Private Bag 4800, Christchurch 8140, New Zealand\\
$^{3}$National Astronomical Observatory of Japan, Mitaka-shi, Tokyo 181-8588, Japan
}

\date{Accepted XXX. Received YYY; in original form ZZZ}

\pubyear{2025}

\begin{document}
\label{firstpage}
\pagerange{\pageref{firstpage}--\pageref{lastpage}}
\maketitle

\begin{abstract}
    %*INTRO
    The super star cluster NGC1569-B has recently been observed to have an extremely high [Ba/Fe].
    We consider that the observed high [Ba/Fe] ($\sim 1.3$) is due to the chemical enrichment of giant molecular clouds by either collapsars, neutron star mergers, or magneto-rotational supernovae, and thereby investigate which of the three polluters can best reproduce the observed [Ba/Fe].
    %*METHOD
    Since it is found that collapsars can best reproduce such an extremely high Ba abundance, we numerically investigate the star cluster formation in NGC1569 using chemodynamical simulations of merging dwarf galaxies with chemical enrichment by collapsars.
    %* Results
    The principal results are as follows.
    First, a cluster of the same scale as NGC1569-B was found to match both the observed [Ba/Fe] and [Fe/H] values, the best cluster having [Ba/Fe]$=1.3\pm0.2$ and [Fe/H] $=-0.7\pm0.2$.
    This simulation used a core-collapse supernova per collapsar rate of 70, a standard initial mass function and an initial metallicity of [Fe/H]=-1.5.
    Second, a prediction of the Eu abundance of NGC1569-B is made: [Eu/Fe]$=1.9\pm0.2$.
    These results are shown to be invariant under a change in the orbit parameters used for the merger.
    The need for a merger to promote the star formation that leads to the synthesis of the Ba and the star cluster formation is confirmed.
    %*Conclusion
    Collapsars can not only better explain [Ba/Fe] but also be consistent with the observed star formation rate and stellar mass of the dwarf galaxy.
\end{abstract}


%%%%%%%%%%%%%%%%%%%%%%%%%%%%%%%%%%%%%%%%%%%%%%%%%%

%%%%%%%%%%%%%%%%% BODY OF PAPER %%%%%%%%%%%%%%%%%%

\section{Introduction}\label{Sec:Intro}

%1. Abundance spreads in CNO, Fe, r- s-. SC formation. Why observations are important
Recent photometric and spectroscopic observations of the Galactic globular clusters (GCs) have revealed that the GCs have internal abundance spreads in light \citep{Osborn1971,Sneden1992,Gratton2004,Melendez2009},  s-process \citep[e.g.][]{Marino2009}, and r-process elements \citep[e.g.][]{Roederer2011}.
These observations also have discovered intriguing anti-correlations between O and Na \citep{Sneden1992,Carretta2009,Carretta2010,Gratton2015},  C and N \citep[e.g.][]{Bell1980},  Mg and Al \citep[e.g.][]{Pancino2017}, and K and Mg \citep{Mucciarelli2012,Carretta2021}, which can possibly provide strong constraints on the theories of GC formation.
These all imply there are multiple chemical populations (MPs) in most GCs, which are not present in young massive clusters (YMCs) \citep{Gratton2012,Bastian2017}.
MPs are important to understand as they are key links to the formation history of the GC due to the different processes that produce each trend.

%2. r- spread is important in constraining SC formation history
Of concern to this work are the rapid neutron capture (\textit{r}-)process elements \citep{Arnould2007,Cowan2021}.
The spread in these \textit{r}-process elements were reviewed across many galactic GCs by \citet{Roederer2011}.
Since then, the cluster M15 has been repeatedly observed. A bimodal spread in Ba (and other \textit{r}-process elements) was found \citep{Worley2013}, and there was no trend in this dispersion with stellar luminosity \citep{Kirby2020}, so the Ba must have been made before or during the cluster formation.
This constrains the timescale of the events used to do the pollution.
M92 had \textit{r}-process dispersion in only the first generation of stars, with the younger generation showing no variation \citep{Kirby2023}, which indicates that the pollution occurs within a short time of the birth of the first generation, and with a delay in the second generation to allow for the mixing of the \textit{r}-process elements.
Combining \textit{s-} and \textit{r-}process Ba has been investigated, to reproduce the abundance parterns in the Galaxy \citep[e.g.][]{Raiteri1999}

%3. NGC1569-B introduction and observations 
Recently a super star cluster (SSC) NGC1569-B has been observed to have an over-abundance of Ba, [Ba/Fe] = $1.28 \pm 0.21$ \defcitealias{Gvozdenko2022}{G22}\citep[hereafter \citetalias{Gvozdenko2022}]{Gvozdenko2022} using an integrated light spectrum.
They also report a [Fe/H] =-0.74 $\pm$ 0.05.
This Ba abundance is much higher than anything measured previously and requires further study to try and recreate a star cluster with [Ba/Fe] so super solar.
The [Ba/Fe] reported in \citetalias{Gvozdenko2022} is seemingly independent of their changing the ratio of red to blue supergiants.
With such an invariant value under such a change and the extremely super solar nature of the Ba abundance, this work uses the largest uncertainty given, $\pm0.21$.
The attempted explanation in \citetalias{Gvozdenko2022} for this high [Ba/Fe] from asymptotic giant branch (AGB) stars synthesising Ba through the slow neutron capture (\textit{s}-)process is unsatisfying, as the authors admit because the AGB pollution timescale is larger than the cluster age.

%4. NGC1569 gal. properties
NGC1569 is a blue compact dwarf (BCD) galaxy \citep{Israel1988}.
There are two SSCs inside NGC1569, labelled -A and -B. This work focuses on NGC1569-B, which has an age of $15-\qty{25}{\mega\year}$ \citep{Larsen2007}.
The star formation history (SFH) of NGC1569 has been a complicated one, \citet{Angeretti2005} concluded that it is most likely that there have been three periods of intense star formation (SF) in the last $1-\qty{2}{\giga\year}$, the youngest of which occurred between 37 and $\qty{13}{\mega\year}$ ago.
This is corroborated by \citet{McQuinn2010}, which has a burst occurring for most of the last $\sim \qty{100}{\mega\year}$, and \citet{Anders2004} found a burst starting around $\qty{25}{\mega\year}$ ago, in agreement with the SSC age.
The outer region of this galaxy has been consistently forming stars for the whole Hubble time, as shown by \citet{Grocholski2012}.
This relatively constant SF is thought to have ceased $\sim \qty{0.5}{\giga\year}$ ago, and the bursts come after this.
The stellar population with an age $>\qty{10}{\giga\year}$ has a considerable [Fe/H] spread from -1 to -2.
This older population has a uniform spatial distribution, unlike concentrated newer stars \citep{Aloisi2001}.
The distance to NGC1569 used in these works was often quoted as $\qty[separate-uncertainty=true]{2.2(0.6)}{\mega\pc}$, citing \citet{Israel1988}.
\citet{Grocholski2012} disagree, based on their earlier work \citep{Grocholski2008}, with new data that resolves the tip of the red giant branch, finding a distance of $\qty[separate-uncertainty=true]{3.06(0.18)}{\mega\pc}$. \citet{Grocholski2008} showed that any measurements of the SF rate (SFR) were too small because of the shorter distance used.
The distance change updates the mass of NGC1569-B from \citet{Larsen2007} to $\qty{6.1e5}{\Msun}$.
\citet{Hunter2004} uses another distance of $\qty{2.5}{\mega\pc}$, citing \citet{OConnell1994}.
The distance corrected value for SFR becomes $\qty{0.48}{\Msun\per\year}$.
This value is used for all further SFR calculations. The SFR surface density, $\Sigma_{\text{SFR}}$, invariant with distance, is tabulated at $\qty{1.29}{\Msun\per\year\per\kilo\pc\squared}$ \citep{Hunter2004}.
A merger or interaction in the NGC1569-B's past is likely, such as the ones described in \citet{Johnson2013} or \citet{Stil1998}.
This merger scenario explains the starburst \citep{Schweizer2005, Barnes2004}.
The SFH is consistent with the general case of a bursty SF period during a merger \citep{Cortijo-Ferrero2017} and the drop in SFR after the merger \citep{Pearson2019}.

%5. Summary of previous models of high r- process
There have been attempts made to simulate the observed dispersion of \textit{r}-process elements in galactic GCs.
\citet{Bekki2017} uses a model where the higher \textit{r}-process abundances are in a second generation of stars, whereas \citet{Tarumi2021} use a single \textit{r}-process polluting event before the cluster forms.
Both of these use neutron star mergers (NSMs) as the polluter.
Also, they use Eu as the tracer of the \textit{r}-process, not Ba, because of the possible \textit{s}-process origin of the Ba.
Eu was not measured in \citetalias{Gvozdenko2022}.
\citet{Tsujimoto2017} measure both Ba and Eu, along with Y, and explain the enrichment of dwarf spheroidal galaxies with both NSMs and magneto-rotational supernovae (MR-SNe).
These models explain very subsolar abundances, unlike the observations of \citetalias{Gvozdenko2022} being investigated here.
All the \textit{r}-process elements in these models are often hypothesised to be made in a single event, due to the low abundances being explained.
Because the [Ba/Fe] of NGC1569-B is so supersolar, all forms of \textit{r}-process Ba pollution need to be considered, as well as multiple events.

Common envelope jet supernova (CEJSN) are another proposed \textit{r}-process site \citep{Grichener2019a}.
These occur when a NS is engulfed by a companion giant star.
CEJSN can reproduce solar abundances for the lanthanides and beyond \citep{Jin2024}, and have also been used to explain the presence of \textit{r}-process elements in the Milky Way \citep{Grichener2022a} and ultra faint dwarf (UFD) galaxies \citep{Grichener2022b}.
While these CEJEN can explain many of the \textit{r}-process observations, They have similar total \textit{r}-process yields to, and rates of, NSMs \citep{Grichener2019b}, so we don't consider them further.

NSMs and MR-SNe are considered here, as well as a different polluter, collapsars (COLs) \citep{MacFadyen1999}.
These are another type of supernova of rotating massive stars that have higher \textit{r}-process yields \citep{Siegel2019}.
\citetalias{Gvozdenko2022} was the first [Ba/Fe] measurement made of NGC1569-B, and this work is the first attempt to model the extremely high value measured.
The different \textit{r}-process yields and time scales of the three pollution sites \citep{Siegel2022} can be used to distinguish which of them is the dominant polluter of this cluster, as the [Ba/Fe] is so high.

%6. Purpose of paper 
The purpose of this paper is to reproduce a cluster with similarly high [Ba/Fe] based on new computer simulations of GC formation, incorporating chemical enrichments of the interstellar medium (ISM) by collapsars in merging between gas-rich dwarf galaxies.
We particularly investigate the mean [Ba/Fe] of stars in the simulated star clusters (SCs) and the internal [Ba/Fe] spread among the SC member stars for each SC.
Although we have already investigated SC formation in interacting \citep[e.g.][]{Bekki2004,Williams2022} and merging galaxies \citep[e.g.][]{Bekki2002}, we did not investigate the Ba abundance properties of the simulated SCs.
Therefore, our new simulations can provide new clues to the origin of SCs with unusually high [Ba/Fe] formed in dwarf-dwarf merging.
Our previous simulations also showed that secondary star formation from gas accreted onto existing SCs is possible \citep[e.g.][]{Bekki2008,McKenzie2018}, which means that [Ba/Fe] of second-generation stars can be quite high if new stars are formed from accreting gas mixed with Ba-rich ejecta from low-mass AGB stars.
However, we do not consider this possibility in this paper at all, because the SCs in NGC1569 are so young that such AGB stars are unlikely to contribute to early chemical enrichment in SC formation.

Dwarf galaxy mergers are sites of high star formation \citep[e.g.][]{Renalud2022}.
Clusters are often formed in simulations of these events \citep[e.g.][]{Lahen2019}.
While it is not without challange \citep{Hislop2022}, the formation of clusters can be done over a large redshift range \citep{Ma2020}.
The phyical properties of groups of star clusters formed from simulated mergers of dwarf galaxies are also studied \citep[e.g.][]{Elgreem2024}.

Although there are several recent simulations on the SC formation and chemical abundance patterns of
the simulated SCs \citep[e.g.][]{Lahen2024},  we do not discuss these results much, because we focus exclusively on the [Ba/Fe] of SCs in BCD galaxies.

%7. outline of paper
This paper is laid out as follows:
The model is presented in \autoref{Sec:Model}, with the three \textit{r}-process polluters discussed in turn in \autoref{SubSec:3PosPol}, the simulations of the best polluter are then described for the rest of the section.
In \autoref{Sec:Res} we present the fiducial model from our simulations and discuss the properties of this chosen cluster and how the results are affected by a change in the parameters of the merger.
These results are further discussed and put in context in \autoref{Sec:Dis}.
The key takeaways and a conclusion are presented in \autoref{Sec:Conc}.

\section{The Model} \label{Sec:Model}

We consider that collapsars, neutron star mergers, and magneto-rotational supernovae could all possibly pollute the ISM of the dwarf galaxy with their ejecta rich in r-process elements in the present study.
Core-collapse supernovae (CCSNe) have long been speculated to be \textit{r}-process sites, along with NSMs, but there is great debate over the contribution for each kind of site \citep[summarised in][]{Siegel2022}.

We first investigate the required numbers of COLs, NSMs and MR-SNe ($N_{\text{COL}}, N_{\text{NSM}}, N_{\text{MR-SN}}$ respectively) that are needed to pollute the ISM to the observed level.
From these values, we then derive the required masses of all stars that need to form for the pollution ($M_{*\text{,COL}}, M_{*\text{,NSM}}, M_{*\text{,MR-SN}}$) for a given set of parameters; the initial mass function (IMF) and event rates.
By doing so, we can investigate (i) whether $M_{*\text{,COL}}$ etc. are possible  for such a small dwarf galaxy like NGC1569 thus (ii) which of the three is the most reasonable polluter for of Ba for this cluster.

Then we perform computer simulations with chemical enrichment by the best polluter and attempt to find a SSC similar NGC1569-B in [Ba/Fe], [Fe/H] and mass.
These simulations vary physical parameters of the system, that can be linked back to observables.
We consider that this two-stage investigation is the best way to understand the origin of rather high [Ba/Fe] that was found by \citetalias{Gvozdenko2022}, because it is extremely time-consuming to investigate all of these three possibilities using numerically
costly computer simulations of dwarf galaxy mergers.

\subsection{Three Possible Polluters} \label{SubSec:3PosPol}

\subsubsection{Collapsars} \label{SubSubSec:Cols}
COLs are a well-studied subset of CCSNe, first postulated as ``failed'' supernova \citep{Woosley1993} because the explosion happens slightly delayed from the core-collapse, often as a (long) gamma-ray burst (GRB) \citep{Woosley2006}.
The Ba that COLs produce is made through the \textit{r}-process.
An accretion disk is formed around the black hole that occurs when the core-collapse marks the death of the COL progenitor.
This disk becomes opaque to the neutrinos it produces, which starts a feedback loop that is a long theorised \citep{Pruet2004,McLaughlin2005,Kohri2005,Siegel2019,Miller2020,Zenati2020,Brauer2021,Barnes2023} site of nucleosynthesis through the \textit{r}-process.
However, recent observations of some supernovae \citep{Anand2024} and a GRB \citep{Blanchard2024} are not finding signatures of the \textit{r}-process, although these signatures should exist \citep{Barnes2022}.
This does throw some doubt on the COL model of \textit{r}-process pollution, but more work is needed in this area.
The accretion disk in a COL also powers the jets and the resulting explosion of a GRB \citep{MacFadyen2001} or jet driven supernova \citep{Heger2003}.
COLs form more heavy \textit{r}-process elements than NSMs by up to 30 times the yield \citep{Siegel2019}, and can be thought of as the dominant polluter of the ISM for these species, especially on short time scales.
A lower metallicity increases the chance of COLs because Wolf-Rayet (WR) stars wind-driven mass loss increases with metallicity \citep{Crowther2007} and some stripping is necessary for a GRB, but an excess leads to lower angular momentum and therefore fewer COLs \citep{Fryer1999}.
The envelope stripping could also be done by a binary companion, with $\sim 70\%$ of O stars in binaries \citep{Sana2012} the chances of a COL occurring increases.
Because their progenitors are massive stars, COLs can reasonably be assumed to happen without a large delay, as their main sequence (MS) lifetime is only a few $\unit{\mega\year}$.

\begin{figure}
    \includegraphics[width=\columnwidth]{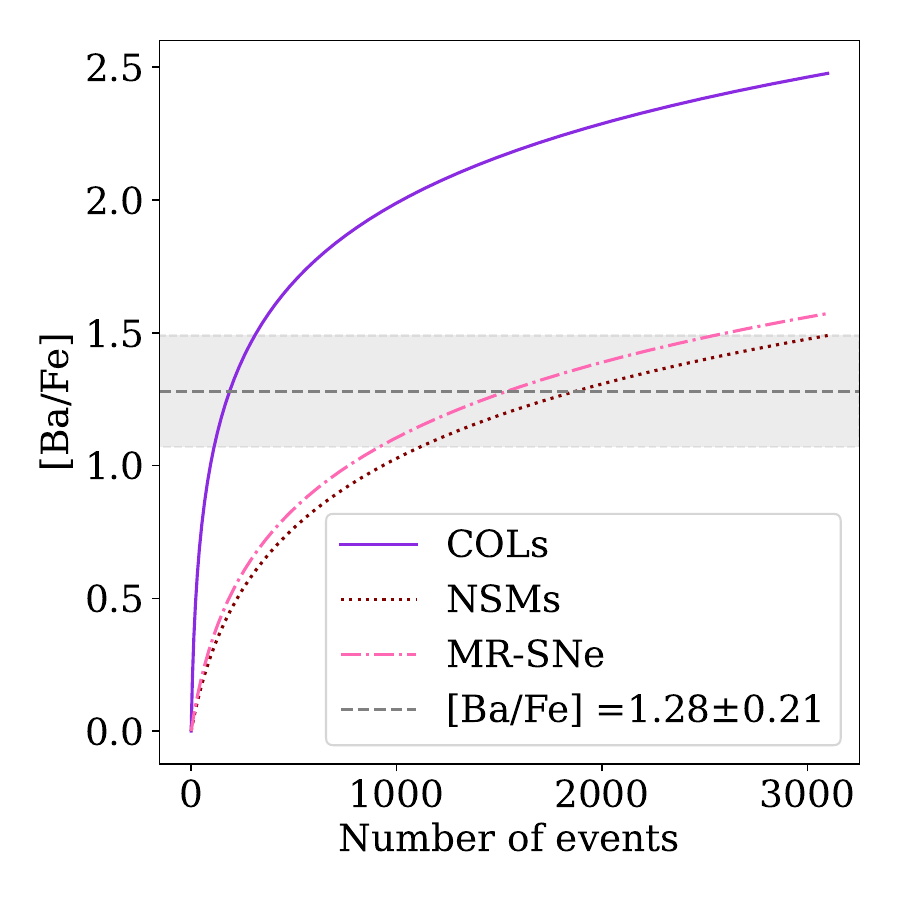}
    \caption{The cumulative [Ba/Fe] for the three different types of polluters increasing with the number of events. COLs (solid purple line) are far more efficient at increasing the Ba abundance than either the NSMs (dotted brown line) or MR-SNe (dot-dashed pink line), due to the higher $m_{\text{ej,Ba}}$. The observed value of [Ba/Fe] in NGC1569-B ($1.28 \pm 0.21$, \citetalias{Gvozdenko2022}) is shown by the dashed grey line for the value and the rectangle for the uncertainty.}
    \label{Fig:ThreePols}

\end{figure}

To calculate the number of COLs, $N_{\text{COL}}$, needed to pollute a GMC to the extreme level of [Ba/Fe] being investigated, the initial metallicity of the system is used to get the initial Ba concentration, and thus an initial mass of Ba, $M_{\text{Ba,i}}$.
This is then incremented by the yield of Ba from a single COL, $m_{\text{ej,Ba,COL}}$, until the [Ba/Fe] matches that of the observation \citetalias{Gvozdenko2022}.
The yields used here are those needed to reproduce galactic abundances, from \citet{Siegel2019}.
The Ba produced in each COL will not all go into the GMC, as this depends on where the COL occurs, the recycling efficiency of the ISM, and many other unknowns.
The fraction of Ba retained by the GMC is denoted $f_{ret}$, the retention parameter.
$f_{ret}=1$ implies all the Ba from every COL goes into the SSC.
This is unphysical as while the ejecta may not escape the galaxy, not all the ejecta from each event will go into the one GMC that eventually forms the SSC, so smaller values of $f_{ret}$ are used.
This leads to the final mass of Ba, $M_{\text{Ba,f}}$, which can be found by
\begin{equation} \label{Eq:BaMass}
    M_{\text{Ba,f}} = M_{\text{Ba,i}} + m_{\text{ej,Ba,COL}} \cdot f_{ret} \cdot N_{\text{COL}}.
\end{equation}
Allowing for all the ways \textit{r}-process COL ejecta could not make it to a GMC with an $f_{ret} = 0.1$, \autoref{Eq:BaMass} can be inverted for the number of COLs.
The solid purple line in \autoref{Fig:ThreePols} shows this, with the cumulative addition of Ba from COLs far outpacing NSMs and MR-SNe.
This has been done under some fairly generous assumptions: COLs don't add Fe to the system, nor do any other CCSN, and the metallicity of the system stays constant at the observed [Fe/H] = -0.74 \citetalias{Gvozdenko2022}.
The rates of the other polluters will get the same generosity.

\begin{equation}\label{Eq:CCSNfromCOL}
    N_{\text{CCSN}} = N_{\text{COL}} \cdot f_{\text{COL}}.
\end{equation}
Multiplying the number of COLs by an appropriate ratio of CCSN per COL, $f_{\text{COL}}$, (a key parameter of the simulations, see \autoref{SubSubSec:SNperCOL} below for a larger discussion of this parameter) gives the number of CCSN needed, $N_{\text{CCSN}}$, as shown in \autoref{Eq:CCSNfromCOL}.
A conservative value of $f_{\text{COL}}=1000$ is used for this calculation.
Integrating over all the high-mass stars that can undergo a CCSN (\qtyrange{8}{50}{\Msun}), the fraction of CCSN per unit solar mass is found by
\begin{equation}\label{Eq:CCSNIMF}
    N_{\text{CCSN,0}} = C_{\text{IMF}} \int_{\qty{8}{\Msun}}^{\qty{50}{\Msun}} \phi(m)\,dm.
\end{equation}
\autoref{Eq:CCSNIMF} uses an initial mass function (IMF), $\phi(m)=m^{-\alpha}$ with slope $\alpha = 2.35$ \citep{Salpeter1955}, with $C_{\text{IMF}}$ being the normalisation constant given by
\begin{equation}\label{Eq:IMFNorm}
    1= C_{\text{IMF}} \int_{\qty{0.1}{\Msun}}^{\qty{50}{\Msun}} \phi(m) m\,dm.
\end{equation}
Other IMFs give slightly different results.
There is a lower SN rate in those with similar high-mass slopes but different normalisations due to lower mass turnoffs, such as the IMF of \citet{Chabrier2003}.
As there are a fewer high mass stars being formed, a larger total mass of star formation is needed to have the same number of SN.

The mass of stars that need to form is given by
\begin{equation}\label{Eq:MassCol}
    M_{*\text{,COL}} = \frac{N_{\text{CCSN}}}{N_{\text{CCSN,0}}}.
\end{equation}
Using an SFR \citep[the $\qty{0.48}{\Msun\per\year}$ from][]{Hunter2004} the time to form this mass of stars, $\Delta t_{\text{SF,COL}}$, is found by
\begin{equation}\label{Eq:TimeFromSFR}
    \Delta t_{\text{SF,COL}} = \frac{M_{*\text{, COL}}}{\text{SFR}}.
\end{equation}

The values are calculated to be $M_{*\text{,COL}}=\qty[separate-uncertainty=true]{2.6(1.7:1.1)e7}{\Msun}$ of stars need to form, which takes $\Delta t_{\text{SF,COL}}=\qty[separate-uncertainty=true]{53(36:22)}{\mega\year}$ by \autoref{Eq:MassCol} and \autoref{Eq:TimeFromSFR}.
The slight dispersion in the lifetimes of these COLs and the SF being an extended event on these timescales means the COLs could all pollute the same GMC, which can live for $\approx \qty{10}{\mega\year}$ \citep{Mouschovias2006}.
This time of star formation fits well with the SFH of NGC1569 \citet{Angeretti2005}, which used higher but less certain SFRs which would only decrease the time of star formation. So the collapsar model makes sense for this pollution of Ba. These values depend on the slope of the IMF chosen, a \citet{Chabrier2003} IMF gives about a factor of 4 increase.
We investigate changing the IMF to an SFR-dependent one during the simulations.

\subsubsection{Neutron Star Mergers} \label{SubSubSec:NSMs}
NSMs were observationally confirmed by their gravitational wave signature \citep{Abbott2017} and have been used to explain the \textit{r}-process pollution for a long time, see the models in other works discussed above for more information.
They have often been invoked as single-event polluters, but the extreme [Ba/Fe] being investigated here requires many NSM events.
We now calculate how many NSMs are expected in NGC1569 from the old stellar population, and the number needed to raise the [Ba/Fe] to the necessary level.

Because of the long time delay, the mass of stars that need to form for the NSM pollution, $M_{*\text{,NSM}}$, cannot be calculated.
However, the number of NSMs that occur at a given time from the old stellar population of NGC1569 in some change in time $\Delta t$ can be calculated.
The NSMs from this population are proportional to a few quantities related to the number of NS binaries.
The binary star fraction, here assumed to be \qty{50}{\percent}, this comes from the \citet{Sana2012} binary fraction of \qty{70}{\percent} and their claim that at least \qty{20}{\percent} of these will result in mergers for the O-type stars while they are still on the main sequence.
These values are also adopted for the early B type stars that can become neutron stars.
The fraction of stars that form neutron stars $N_{\text{NS,0}}$, here taken to be the number of stars made between \qtyrange{8}{20}{\Msun}, is also important.
\begin{equation}\label{Eq:NSMIMF}
    N_{\text{NS,0}} = C_{\text{IMF}} \int_{\qty{8}{\Msun}}^{\qty{20}{\Msun}} \phi(m)\,dm
\end{equation}
This uses the same $C_{\text{IMF}}$ as found in \autoref{Eq:IMFNorm}, and calculates the fraction of stars in that mass range, so \autoref{Eq:NSMIMF} is much like \autoref{Eq:CCSNIMF}.
Also normalising an NS merger function
\begin{equation}\label{Eq:NSMNorm}
    1 = C_{\text{NSM}} \int_{\qty{0}{\giga\year}}^{\qty{13}{\giga\year}}\frac{t_{min}}{t}\,dt
\end{equation}
to solve for $C_{\text{NSM}}$ as the normalisation constant, with $t_{min}=\qty{700}{\mega\year}$ as the minimum delay time of an NSM.
Having all the mergers occur during the lifetime of the galaxy so far is generous because there is no reason they should have all occurred already.
Integrating this same function as in \autoref{Eq:NSMNorm} over the $\Delta t=\qty{10}{\mega\year}$ lifetime of a GMC \citep{Mouschovias2006} at the galaxies' current age  \citep[\qty{13}{\giga\year},][]{Grocholski2012}

\begin{equation}\label{Eq:NSMpMsun}
    N_{\text{NSM,0}}= 0.5 \cdot C_{\text{NSM}} \cdot N_{\text{NS,0}}\int_{\qty{13}{\giga\year}}^{\qty{13.01}{\giga\year}}\frac{t_{min}}{t}\,dt
\end{equation}
gives a number of NSM per solar mass in this $\Delta t$, where the 0.5 coefficient comes from the \qty{50}{\percent} binary fraction.
Multiplying this value by the mass of stars in the galaxy, $M_{*} = \qty{2.8e8}{\Msun}$ \citep{Johnson2012} gives 84 NS mergers occurring in NGC1569 when the SSC was forming.
This is galaxy-wide, so not all the \textit{r}-process Ba synthesised by these events will be incorporated into the GMC.

The number of NS mergers needed to reach the high [Ba/Fe] measured by \citetalias{Gvozdenko2022} can be calculated similarly to the number of COLs needed.
By replacing the \textsubscript{COL} values with \textsubscript{NSM} parameters in \autoref{Eq:BaMass}, the $N_{\text{NSM}}$ value can be calculated.
Once again an $f_{ret}$ is applied to account for all the ways that the Ba does not make it to the GMC.
For consistency, $f_{ret} = 0.1$ has been used for NSMs as well.
The [Ba/Fe] increase with $N_{\text{NSM}}$ is shown in \autoref{Fig:ThreePols} for NSMs (in dotted brown) too.
It is no accident that an order of magnitude more NSMs are needed over the COLs.
The yields used have been $\times10$ less than the COL yield, as suggested by \citet{Siegel2019}, (i.e. $m_{\text{ej,Ba,NSM}} = 0.1 \cdot m_{\text{ej,Ba,COL}}$).
The $N_{\text{NSM}}$ estimated from \autoref{Fig:ThreePols} is clearly more than the 84 NSMs that occur in $\Delta t = \qty{10}{\mega\year}$ from the \qty{13}{\giga\year} age of this population.
This same mass of ``old'' stars would need to be only \qty{600}{\mega\year} old for NSMs to be viable polluters (setting this as the lower bound of integration in \autoref{Eq:NSMpMsun} with the same $\Delta t$), for this yield and $f_{ret}$ combination.
The mass of stars formed in the bursty star formation in the last \qty{}{Gyr} is too small, by an order of magnitude, to apply the same argument of NSM pollution from the younger stars.

\subsubsection{Magneto-Rotational Supernovae} \label{SubSubSec:MRSNe}

A similar analysis to the COLs and NSMs can be done for magneto-rotational supernovae. This calculation is shown in \autoref{Fig:ThreePols} with the dot-dashed pink line for MR-SNe. The Ba ejected per event, $m_{\text{ej,Ba,MR-SN}}$, was calculated by \citet{Nishimura2017} \citep[but see also][]{Tsujimoto2017} and used for the substitution into the MR-SNe equivalent of \autoref{Eq:BaMass}.
Once again, $f_{ret}=0.1$ is used, so the results are comparable with the other polluters.

The MR-SNe are closer to the NSMs than the COLs for the number needed, as inferred from \autoref{Fig:ThreePols}.
$N_{\text{CCSN}}$ can be calculated from a parallel to \autoref{Eq:CCSNfromCOL}, the value of $N_{\text{MR-SN}}$ and a CCSN per MR-SN value of $f_{\text{MR-SN}}=200$ \citep{Tsujimoto2015}.
With this value and the same $N_{\text{CCSN,0}}$ from \autoref{Eq:CCSNIMF}, $M_{*\text{,MR-SN}}$ can be calculated in an analogous way to \autoref{Eq:MassCol}.
This implies $M_{*\text{,MR-SN}}=\qty[separate-uncertainty=true]{4.3(2.8:1.8)e7}{\Msun}$, which would be an SF time of $\Delta t_{\text{SF, MR-SN}}=\qty[separate-uncertainty=true]{90(58:38)}{\mega\year}$ from an adjustment of \autoref{Eq:TimeFromSFR}.
Even with MR-SN being a higher fraction of all CCSN than the COL fraction estimate, it is unlikely that these events will be numerous enough for the pollution of Ba in NGC1569-B, as the SF time and mass of stars needed is not in agreement with the observations.
There is also debate over the \textit{r}-process production rates in MR-SNe, with \citet{Mosta2018} arguing that the magnetic fields need to be larger than is realistic, else the \textit{r}-process production is reduced by two orders of magnitude above the \nth{2} peak.

\subsubsection{The Best Polluter}\label{SubSubSec:BestPol}

The estimates in \autoref{Fig:ThreePols} were calculated assuming no Fe was produced in these events.
This was used because the mass of Fe produced in these events are uncertain.
The ratio of each event rate to CCSNe rates is also uncertain, compounding the problem.
To reduce confusion, we opted to not account for increases in Fe, allowing the forthcoming simulations to handle that better than these estimates could have.
However, experimenting with a range of Fe yields from only the Ba pollution events, the results do not change, as the total mass of Fe produced is insignificant.
Adding extra Fe from concurrent SN, COLs were the only polluter to reach the required [Ba/Fe] in a reasonable number of events for any of the scenarios.

From the above analysis, nicely captured in \autoref{Fig:ThreePols}, COLs are the only polluter that could have raised the progenitor GMC of NGC1569-B to a [Ba/Fe] as high as it has been observed \citepalias{Gvozdenko2022}.
The mass of new stars and the time taken to form them are broadly consistent with the SFH in the COL scenario.
MR-SNe are close to being able to reproduce the measurements, but the SF time is too long.
There are not enough NSMs occurring due to the old stellar population to provide the needed NSMs for pollution of [Ba/Fe] up to $\sim 1.3$.

Thus, collapsars are used as the polluter for the simulations of the SSC formation described below.

\subsection{Initial disk models}\label{SubSec:IDiskMod}

Since our previous simulations demonstrated that BCDs like the host dwarf galaxy NGC 1569 can be formed
from mergers between bulgeless and gas-rich dwarf disk galaxies \citep{Bekki2008a}, we here adopt the gas-rich
disk galaxy models for merger progenitor galaxies.  We here briefly describe the disk models,
given that the details of the disk models have been described in our previous papers
\citep{Bekki2008a,Bekki2014}.
A dwarf galaxy is assumed to  consist of dark matter, stars, and gas (without stellar bulge).
The total masses of dark matter halo, stellar disk, and gas disk in a dwarf galaxy are denoted as $M_{\rm dm}$, $M_{\rm s}$, $M_{\rm g}$, respectively.
In order to describe the radial density profile of the dark matter halo of the dwarf galaxy, we adopt the density distribution of the NFW halo \citep{Navarro1996} suggested from CDM simulations:
\begin{equation}
    {\rho}(r)=\frac{\rho_{0}}{(r/r_{\rm s})(1+r/r_{\rm s})^2},
\end{equation}
where  $r$, $\rho_{0}$, and $r_{\rm s}$ are
the spherical radius,  the characteristic  density of a dark halo,  and the scale length of the halo, respectively,
and the reasonable value of the central concentration parameter, $c$
($c=r_{\rm vir}/r_{\rm s}$, where $r_{\rm vir}$ is the virial, predicted from NFW)
is given to each low-mass dark matter halo in the present study (i.e., larger $c$
for larger $M_{\rm dm}$).
We mainly investigate the models with $M_{\rm dm}=\qty{e11}{\Msun}$, $R_{\rm vir}= \qty{45}{\kilo\pc}$, $c=16$, $M_{\rm s}=\qty{6e8}{\Msun}$, and $M_{\rm g}=\qty{1.2e9}{\Msun}$, which is a reasonable set of parameters for the dwarf galaxy  NGC 1569.

The radial ($R$) and vertical ($Z$) density profiles of an initially thin stellar disk with a disk size $R_{\rm s}$ are proportional to $\exp (-R/R_{0}) $ with scale length $R_{0} = 0.2R_{\rm s}$  and to ${\rm sech}^2 (Z/Z_{0})$ with scale length $Z_{0} = 0.04R_{\rm s}$, respectively, in all dwarf galaxy models.
The gas disk with a size  $R_{\rm g}=2R_{\rm s}$
is also assumed to have a scale length of $R_{0, g}$ and
a vertical scale length $Z_{0,g}$.
In addition to the rotational velocity caused by the gravitational field of stellar and gaseous disk, bulge, and dark halo components, the initial radial and azimuthal velocity dispersions are assigned to the disc component according to the epicyclic theory with Toomre's parameter $Q$ = 1.5 for all dwarf galaxy models.
The vertical velocity dispersion at a given radius is set to be 0.5 times as large as the radial velocity dispersion at that point.
We investigate only  the thin disk  models with $R_{\rm s}=5$ kpc (i.e., $R_0=\qty{1}{\kilo\pc}$) in the present study, though the results might depend on the initial disk structures
and kinematics.
A stellar bar, which can possibly trigger a central starburst, cannot be formed from gravitational instability in the initial disks, because the initial stellar disk fractions are  very small ($ \approx 0.006$, i.e., very weakly self-gravitating).

In order to investigate chemical enrichment, star formation, and dust formation and evolution in dwarf-dwarf mergers, we adopt the same code and model as those used in our previous studies that investigated metal and dust enrichments in galaxies
\defcitealias{Bekki2013}{B13} \citep[hereafter \citetalias{Bekki2013}, \citet{Bekki2015}]{Bekki2013}.
Since the details are given in \citetalias{Bekki2013}, we here briefly describe the models.
The only minor difference between this study and \citetalias{Bekki2013} is that the time evolution of Ba and Eu in galaxies  is newly investigated in the present chemodynamical simulations: it should be stressed here that dust-phase Ba and Eu abundances are not considered in the present study due to the lack of knowledge about their depletion levels in ISM.
The mass and size resolutions of the simulations presented here are \qty{2000}{\Msun} and \qty{35}{\pc},
respectively, for $M_{\text{dm}}=\qty{e10}{\Msun}$ models.
This resolution is limited by the available compute time for the project.
Galaxy-wide star formation in dwarf mergers is  based on the observed  Kennicutt-Schmidt law, i.e., SFR$\propto \rho_{\rm g}^{\alpha_{\rm sf}}$ \citep{Kennicutt1998}, where $\alpha_{\rm sf}$ is the slope of the power-law and set to be 1.5 in the present study, as observed.
Conversion from gas particles into new collisionless stellar particles is assumed to occur if the local density exceeds a threshold density for star formation ($\rho_{\rm th}$).
As described later, this threshold gas density can have an influence on the final [Ba/Fe] of the simulated SCs. Feedback effects from CCSNe and SNIa are both incorporated, though those from NSMs are yet to be done. The same chemical yields from CCSNe, SNIa, and AGB stars as those used in \citetalias{Bekki2013} are adopted in the present study for the Salpeter IMF \citep{Salpeter1955}.
The model parameters for dust formation, growth, and destruction (e.g., dust yields and growth timescale) adopted in the present study are exactly the same as those in \citetalias{Bekki2013}.

The chemical evolution is treated robustly in these simulations.
We assume Fe is formed in CCSNe and prompt type Ia SN.
The number of massive stars are assigned to a star particle accor to a \citet{Salpeter1955} IMF.
The metals are formed according to the yeilds of \citet{Tsujimoto1995}, and the main-sequence lifetimes of the projenitor.
Type Ia SN start their explosions after just \qty{100}{\mega\year}.
The \textit{r}-process elements are formed in COLs, as they were shown to be the best polluter above (\autoref{SubSec:3PosPol}), and can star occuring \qty{10}{\mega\year} after SF.

Initial [Fe/H] are assumed to be free parameters so that the best initial [Fe/H] can be derived to explain the observed final [Ba/Fe] and [Fe/H] of NGC1569-B.

\subsection{Orbital parameters for mergers}\label{SubSec:OrbitParam}

The orbit of a dwarf-dwarf merger is assumed to be parabolic with an orbital eccentricity of 1.0
in all models.
The apocenter and pericenter ($R_{\rm p}$) distances are set to be $10R_{\rm s}$ (i.e., ten times the stellar disk size) and $0.1R_{\rm s}$, respectively, in most models.
The mass-ratio of two dwarfs represented by $m_2$ is a free parameter that can influence the strength of a starburst and the associated SC formation.
We mainly investigate the major merger models with $m_2=1$.
The orbit of the two dwarfs is set to be the $xy$
plane and the spin of each galaxy in a merger is specified by two angles $\theta_i$ and $\phi_i$, where suffix i (=1 and 2, i.e., primary and companion galaxies)  is used to identify each galaxy. $\theta_i$ (in units of degrees) is the angle between the $z$-axis in the adopted coordinate system and the vector of the angular momentum of a disc.
$\phi_i$  is the azimuthal angle measured from the $x$-axis to the projection of the angular momentum vector of a disc onto the $xy$ plane (i.e., orbital plane).
We mainly show the results of a major merger model  with
$\theta_1=30$,
$\theta_2=45$,
$\phi_1=30$,
and $\phi_2=120$ in the present study, though the orbital configurations can slightly influence SC formation processes.

\subsection{Parameters for Collapsars}\label{SubSec:ColParams}
It is clear from the above analytical tests that the COLs are the only viable way of polluting Ba to the needed levels.
So the simulations below use COLs as the polluter of \textit{r}-process elements.
Here we describe some parameters of the simulations that get changed in our suite of models.

\subsubsection{CCSN per COL} \label{SubSubSec:SNperCOL}
There is a lot of debate around the rates of COLs.
Converting the values given in other works to the simulation input value of CCSN per COL, how many massive stars there are before one of them is a collapsar, gives a wide range this parameter could take.
\citet{Siegel2019} and references therein give one way to calculate a value for this parameter based on the current estimates of GRBs.
A long GRB rate of $\qty[separate-uncertainty=true]{1.3(0.6:0.7)}{\per\giga\pc\cubed\per\year}$ \citep{Wanderman2010}, a GRB beaming fraction $f_b\sim\qty{5e-3}{}$ from an average opening angle $\theta_j=\qty{6}{\degree}$ \citep{Goldstein2016}, which gives a COL rate of $\sim\qty{260}{\per\giga\pc\cubed\per\year}$.
With a CCSNe rate of $\qty[separate-uncertainty=true]{7.05(1.43:1.25)e4}{\per\giga\pc\cubed\per\year}$ \citep{Li2011}, the CCSN per COL is estimated to be $\sim 270$.
Using this as the $f_{\text{COL}}$ value in the above calculations would have lowered the mass of new stars and the time of SF for COLs by a factor of $\sim4$.
Earlier work, \citet{Paczynski1998}, has a CCSN per COL of $10^4-10^5$. These values are just using observed GRB rates, so are underestimates, as not all COLs will be GRBs.
A jet-driven supernova could be the result of a COL if the H envelope remained \citep{Heger2003} so GRB rates are only a bound on the COL rates.
\citet{MacFadyen1999} has a conservative collapsar estimate of $<\qty{1}{\percent}$ of CCSNe, so the parameter would be inverted, $>100$.
At a lower metallicity, \citet{Heger2003}  predict GRBs and jet-driven SNe are up to $\sim \qty{10}{\percent}$ of CCSN, which would make the CCSN per COL as low as 10.
The per-galaxy rates of \citet{Fryer1999} have been converted to a CCSN per COL value of between 20-2000.
Inverting the $f_r$ of \citet{Brauer2021}, they take a fiducial value for this parameter at $10^3$, but investigates a range from $10 - 10000$.
These sources give a range of $\log(\text{CCSN per COL})$ from  1 to 5, which is a large spread, explaining why $f_{\text{COL}}=1000$ was a conservative estimate.
In a lower metallicity environment like NGC1569 the lower values for CCSN per COL are acceptable.
\citetalias{Gvozdenko2022} measures [Fe/H]=-0.74 for the B cluster presently, and the old stellar population has $-2 <\text{[Fe/H]}<-1$ \citep{Grocholski2012}.
This low initial metallicity would increase the likelihood of a COL as discussed above, hence a lower CCSN per COL.
This value controls the ratio of \textit{r}-process elements formed in the simulations to the other metals formed in regular CCSNe.

\subsubsection{CCSN per \si\Msun} \label{SubSubSec:SNperMass}
The CCSN per Mass value is dependent on the IMF slope $\alpha$.
This is because a lower $\alpha$ means more high-mass stars are formed, which inherently increases the number of COLs.
A standard IMF is that of \citet{Salpeter1955}.
This has an $\alpha = 2.35$, which gives a CCSN per $\unit{\Msun} = \qty{7e-3}{}$ (\autoref{Eq:CCSNIMF}).

The value of $\alpha$ has been shown to vary with SFR.
In particular, \citet{Gunawardhana2011} show
\begin{align}\label{Eq:AlpSSFR}
    \centering
    -\alpha \approx 0.3 \log(\Sigma_{\text{SFR}}) - 1.7
\end{align}
where $\alpha$ is the IMF power law exponent and $\Sigma_{\text{SFR}}$ is the surface density SFR, as above. \autoref{Eq:AlpSSFR} is a sign changed version of equation (14) from \citet{Gunawardhana2011}, to agree with the sign convention used here of a positive $\alpha$.
Using the $\Sigma_{\text{SFR}}$ from \citet{Hunter2004} for NGC1569 of $\Sigma_{\text{SFR}} = \qty{1.29}{\Msun\per\year\per\kilo\pc\squared}$, \autoref{Eq:AlpSSFR} implies $\alpha = 1.67$.
This $\alpha$ gives CCSN per $\unit{\Msun} = \qty{2.7e-2}{}$.
Changing the IMF is less desirable physically, as \citet{Salpeter1955} was so seminal.

\subsubsection{Other parameters} \label{SubSubSec:OtherParams}
While matching the [Ba/Fe] is important, ignoring all the other outputs of the simulations would invalidate the results.
Importantly the metallicity, in [Fe/H].
Before attempting to match the observed [Fe/H], the initial [Fe/H] was the observed value.
This did increase with time in the simulations because CCSNe added Fe to the system.
Matching the output Fe abundance to the observed value was achieved by changing the initial metallicity value ([Fe/H]\textsubscript{i}) of the merging dwarf galaxies.
\citet{Grocholski2012} showed that the old stellar population in NGC1569 has a large metallicity spread in the range of $-2 <\text{[Fe/H]}<-1$, peaking around -1.25.
So initial values from of [Fe/H]\textsubscript{i} = -1.2 and below were tried to find a value that produced a cluster that matched both the observed [Ba/Fe] and [Fe/H].
By lowering this value, the CCSNe can pollute the [Fe/H] to the level observed, $-0.74\pm 0.05$ from \citetalias{Gvozdenko2022}, while the COLs do the Ba pollution.
This was done in the later models, as trying to match two parameters at once would have been too difficult.

To give some indication of the reliability of the results presented below, the parameters of the dwarf galaxy merger can be changed.
The pericentre of the orbit, standard at $\qty{8.75}{\kilo\pc}$, was varied within about a factor of two.
The mass ratio of the two dwarf galaxies, standard at 1:1, was decreased to 1:0.5 for some models.
The change in total mass that this causes is not expected to be an issue.
Combining changes in both parameters is also explored.
The need for a merger is tested by simulating a galaxy in isolation, with otherwise optimal parameters, to see how the SF and \textit{r}-process pollution is affected.

\subsection{Selection of Star Clusters}\label{SubSec:ClusterSelection}
The new star particles that form can be grouped into clusters after the simulation is complete.
This is achieved by chaining nearby star particles together until they compose a grouping on the scale of a star cluster.
These star particles have variable masses, yet they are all lighter than the gas particle that formed them.
Two parameters control the selection of these clusters.
The first is a neighbouring radius (NR), which groups stars together if the star particles are all within NR of the centre of mass (COM). The standard value for this is \qty{35}{\pc}.
The other parameter is a threshold identification mass, the group of star particles must have a mass of at least this value within the NR to be a cluster.
This is standard at $\qty{5e5}{\Msun}$, about the stellar mass of NGC1569-B, which \citet{Larsen2007} put at $\qty{6.1e5}{\Msun}$.
NR is only changed if no clusters are detected at the standard value, this is noted in the last column of the results in \autoref{Tab:Results}.
The mass parameter is never changed.
Due to the large mass requirement for these clusters, we have no problem resolving them with our mass resolution.
However the spatial resolution is the lowest it could be for such identification.

Some groups of stars the simulation identifies as star clusters have intriguing spatial distributions.
There are groups identified with no stars close to the COM of the cluster.
Real clusters often have more stars in their centres, so this is not physical.

A distinguishing cut can be made using the potential of the star clusters.
The linear regression best-fitting line of radius ($R = \sqrt{\bm{x}^2+\bm{y}^2+\bm{z}^2}$) against the measured potential must be negative for a grouping to be a cluster.
This forces there to be more stars in the centre.
Any non-negative potential gradient is classed here as an overdensity of stars.

Each star particle in the simulation has a known abundance.
Taking the arithmetic mean of star particles in a cluster, the [Ba/Fe] can be compared to that observed in \citetalias{Gvozdenko2022} to see if that simulation's set of parameters reproduces a cluster with the super-solar value.
This average value is an appropriate way of making the comparison as \citetalias{Gvozdenko2022} has measured the [Ba/Fe] from an integrated light spectrum, which has the effect of averaging out the contribution from each star in the cluster.
If another way of measuring the Ba abundance was used, i.e. the star with the maximum [Ba/Fe] was taken in each cluster, these results would not be comparable.
When a simulation model forms multiple clusters, the one with the highest average [Ba/Fe] is presented as the result.
There is no stellar [Ba/Fe] distribution observed in NGC1569-B as individual stars are not resolved in the spectrum.
Matching the average value may lead to an understanding, or at least an estimate of, the underlying properties of such a distribution, as the simulations happily provides the data of the abundances for each star particle.

\subsection{Parameter Study} \label{ParamStudy}

\begin{table*}
    \caption{A summary highlighting the process done when changing the parameters of the simulation.
        The key inputs of CCSN per $\unit{\Msun}$, CCSN per COL and initial metallicity ([Fe/H]\textsubscript{i}) and shown in columns 2,3 and 5 respectively.
        The corresponding [Ba/Fe] and final metallicity ([Fe/H]\textsubscript{f}) are given in columns 4 and 6.
        The first column is a model number identifier, and the last column notes if any other parameters have been changed}
    \label{Tab:Results}
    \noindent\makebox[\textwidth]{
        \begin{filecontents*}{Table1.dat}
a & b & c & d & e & f & g    
M1    & 7$\times 10^{-3}$    & 300           & 0.3 $\pm$ 0.1              &    -0.74    & -0.19 $\pm$ 0.09 &   %oz1
M2    & 7$\times 10^{-3}$    & 100           & 0.6 $\pm$0.3               & -0.74  & -0.31 $\pm$ 0.06  &    %oz2
M3    & 7$\times 10^{-3}$    & 100           & 0.4$\pm$0.4          &-0.74 & -0.3 $\pm$ 0.1    & low $\rho_{th}$\textsuperscript{a}, NR\textsuperscript{b} \qty{52.5}{\pc} %oz3
M4   & 7$\times 10^{-3}$    & 30            & 1.3$\pm$0.2                & -0.74 & $-0.22\pm$ 0.09& %oz10
M5   & 2.7$\times 10^{-2}$  & 100           & 1.3$\pm$0.3                & -0.74 &-0.27 $\pm$ 0.08 & %oz15
M6          & 7$\times 10^{-3}$    & 40           & 1.2$\pm$0.4   & -1.2                    & -0.6$\pm$0.1            &                               %oz18 
M7          & 2.7$\times 10^{-2}$  & 150          & 1.5$\pm$0.3   & -1.3                    & -0.6$\pm$0.2            &                               %oz19 
M8\textsuperscript{c}          & 7$\times 10^{-3}$    & 40           & 1.5$\pm$0.3   & -1.3                    & -0.6$\pm$0.2            &   %oz20                
M9          & 7$\times 10^{-3}$    & 30           & -- \textsuperscript{d}         & -1.4                    &            -- \textsuperscript{d}             & ISO \textsuperscript{e}       %oz26                    
M10          & 7$\times 10^{-3}$    & 30           & 1.6$\pm$0.2   & -1.4                    & -0.6$\pm$0.2            &     %oz27    
M11          & 7$\times 10^{-3}$    & 30           & 1.4 $\pm$ 0.3 & -1.5                    & -0.7$\pm$0.1            &        %run oz30, normal from here                        
M12          & 7$\times 10^{-3}$    & 50           & 1.4 $\pm$ 0.3 & -1.5                    & -0.6$\pm$0.2            &                                          
M13 & 7$\times 10^{-3}$    & 70           & 1.3$\pm$0.2   & -1.5                    & -0.7$\pm$0.2            &                         
M14          & 7$\times 10^{-3}$    & 70           & 1.3$\pm$0.2   & -1.5                    & -0.5$\pm$0.2            & low $\rho_{th}$\textsuperscript{a}, NR \qty{52.5}{\pc} \textsuperscript{b}   
M15          & 7$\times 10^{-3}$    & 70           & 1.3$\pm$0.3   & -1.5                    & -0.8$\pm$0.2            & mass ratio\textsuperscript{f} 1:0.5                
M16          & 7$\times 10^{-3}$    & 70           & -- \textsuperscript{d}         & -1.5                    &            -- \textsuperscript{d}             & ISO\textsuperscript{e}, NR\textsuperscript{b} \qty{157.5}{\pc}                
M17          & 7$\times 10^{-3}$    & 70           & 1.4$\pm$0.2   & -1.5                    & -0.7$\pm$ 0.2           & pericentre\textsuperscript{g} \qty{3.5}{\kilo\pc}                  
M18          & 7$\times 10^{-3}$    & 70           & 1.4$\pm$0.3   & -1.5                    & -0.6 $\pm$ 0.2          & pericentre\textsuperscript{g} \qty{17.5}{\kilo\pc}                   
M19          & 7$\times 10^{-3}$    & 70           & -- \textsuperscript{d}          & -1.5                    &       -- \textsuperscript{d}                  & pericentre\textsuperscript{g} \qty{3.5}{\kilo\pc},  mass ratio\textsuperscript{f} 1:0.5 
M20          & 7$\times 10^{-3}$    & 70           &      1.3 $\pm$ 0.2         & -1.5                    &         -0.8$\pm$0.1                & pericentre\textsuperscript{g} \qty{17.5}{\kilo\pc},  mass ratio\textsuperscript{f} 1:0.5 
M21          & 7$\times 10^{-3}$    & 70           &       1.3$\pm$0.3        & -1.5                    &      -0.7$\pm$0.2                   & pericentre\textsuperscript{g} \qty{6.125}{\kilo\pc}                   
M22          & 7$\times 10^{-3}$    & 70           &      1.2$\pm$0.5         & -1.5                    & -0.7$\pm$0.2                        & pericentre\textsuperscript{g}\qty{13.125}{\kilo\pc}
M23\textsuperscript{c}          & 2.7$\times 10^{-2}$    & 270           &     1.3$\pm$0.2   & -1.5                    & -0.7$\pm$0.2            &                      
\end{filecontents*}
\pgfplotstabletypeset[
            col sep=&,
            string type,
            columns/a/.style={column name=Model},
            columns/b/.style={column name=CCSN per $\unit{\Msun}$},
            columns/c/.style={column name=CCSN per COL},
            columns/d/.style={column name=[Ba/Fe]},
            columns/e/.style={column name=[Fe/H]\textsubscript{i}},
            columns/f/.style={column name=[Fe/H]\textsubscript{f}},
            columns/g/.style={column name=Other Changes}]{Table1.dat}
    }
    \begin{flushleft}
        \footnotesize{
            \textsuperscript{a} Low $\rho_{th}$ means the threshold density for star formation in the simulation was decreased from a standard \qty{1000}{\per\cm\cubed} to \qty{100}{\per\cm\cubed}.
            \textsuperscript{b} NR for neighbouring radius. The distance checked for membership to a cluster, standard at \qty{35}{\pc}.
            \textsuperscript{c} These models output the same data as an earlier model, due to the degeneracy in the parameters CCSN per $\unit{\Msun}$ and CCSN per COL. The same output for different input parameters highlights the degeneracy well. M8 had the same output as M7, and M23 had the same output as the fiducial model M13.
            \textsuperscript{d} -- means the (all the) group(s) of new star particles found had a non-negative potential gradient, and thus failed the check for being a cluster. With no clusters forming, there is nothing to measure the [Ba/Fe] of.
            \textsuperscript{e} ISO means there was intentionally no merger in the simulation, the galaxy was isolated.
            \textsuperscript{f} Ratio of the masses of the two galaxies that merge, standard at 1:1.
            \textsuperscript{g} The pericentre distance of the merger orbit, standard at \qty{8.75}{\kilo\pc}.
        }
    \end{flushleft}

\end{table*}

\autoref{Tab:Results} shows a selection of the simulations performed  by this study.
In total, we performed  94 simulations, with input values more extreme than shown here, and equally extreme outputs.
Those shown in this work have had their model number reindexed to start from M1.
The first five simulations shown are representative of those made where the initial [Fe/H] was the value observed in \citetalias{Gvozdenko2022}.
Comparing M2 with M3 and M5, where the CCSN per COL value was held at 100 and the CCSN per $\unit{\Msun}$ and $\rho_{th}$ were changed, respectively.
The [Ba/Fe] changes drastically between these models, so all of these parameters are important.
Both M4 and M5 match the observed [Ba/Fe], with completely different parameters.
The degeneracy between the two main COL inputs is shown here.
This is explained by a lurking variable, a variable not input to the simulations as a parameter but controls the output more than any of the parameters that are.
The COLs per $\unit{\Msun}$ value of the simulation is what is important to the results, but is not a direct input.
It can be derived by dividing the value of CCSN per $\unit{\Msun}$ by the CCSN per COL value.
This lurking variable also explains the same outputs in both M7 \& M8 and M13 \& M23.
Models M6 through M12 show a slow decrease in the initial [Fe/H] values, and the corresponding decrease in the final values.

No matches to both the observed [Ba/Fe] and [Fe/H] values are found until M13.
This is the model that has best matched the observed parameters of NGC1569-B.
The average [Ba/Fe] and [Fe/H]\textsubscript{f} match the values observed by \citetalias{Gvozdenko2022}.
This model had the standard orbital , merger and clustering parameters, threshold gas density $\rho_{th} = \qty{1000}{\per\cm\cubed}$ and Ba yield from COLs $\text{Ba/COL} = \qty{2.3e-3}{\Msun}$.
It uses the \citet{Salpeter1955} IMF CCSN per COL of $7\times 10^{-3}$.
The initial Fe abundance of [Fe/H]\textsubscript{i} $= -1.5$ with a CCSN per COL value of 70 was able to match those observations.
This initial [Fe/H] is in good agreement with the metallicity of the older star population in NGC1569 \citet{Grocholski2012}.
This is the model that will be presented for the rest of this work as the fiducial model.

\section{Results}\label{Sec:Res}

\subsection{The Fiducial Model}\label{SubSec:FidModel}
\begin{figure*}
    \includegraphics[width=\textwidth]{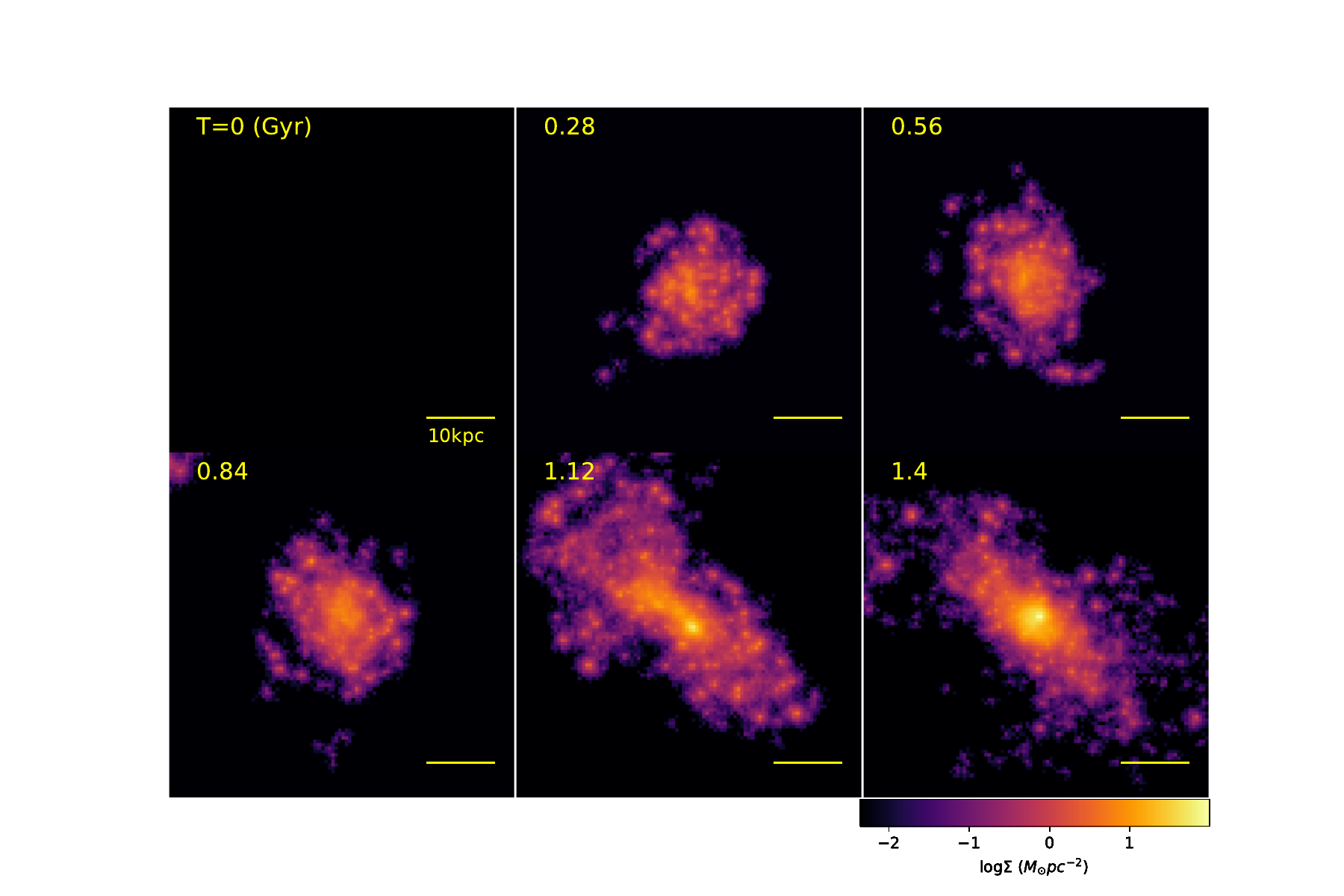}
    \caption{The surface density of new star particles in a log scale, measured in $\unit{\Msun\per\pc\squared}$, of the fiducial simulation.
        The six time steps increase by \qty{0.28}{\giga\year} from the start time at T=0 in the upper left to T=\qty{1.4}{\giga\year} lower right.
        This is centred on one of the dwarf galaxies that are merging.
        A \qty{10}{\kilo\pc} scale bar is provided in each frame.}
    \label{Fig:NewStarDensity}
\end{figure*}

\autoref{Fig:NewStarDensity} describes how numerous young star clusters are formed during the dissipative merging between gas-rich dwarf galaxies in the fiducial model.  After the first pericenter passage, the two start to merge to form a single dwarf galaxy with a strong starburst, merging  about \qty{1}{\giga\year} into the simulation (between pannels 4 and 5).
Due to the adopted high gas fractions in the merging of two dwarf galaxies, high-density gaseous regions corresponding to giant molecular clouds can be quickly formed where star formation can proceed very efficiently.
Compact stellar systems are consequently formed from the new gas clouds, and some of them can become gravitationally bound star clusters.
This effect provides a structure to the merged galaxy and is seen in \autoref{Fig:NewStarDensity} as small regions of high density where the SF has occured.
Many of these clumps cannot stay bound as clusters because of SN feedback.

Five star clusters are identified in the M13 simulation run, they all lie within the NR of \qty{35}{\pc}, have total stellar masses of more than \qty{5e5}{\Msun}, and have a negative potential gradient.
Two more overdensities are identified, which pass the first two checks, but have a positive potential gradient.
The cluster with the highest [Ba/Fe] is presented as the best cluster, with a [Ba/Fe]$=1.3\pm0.2$ and a [Fe/H]\textsubscript{i}$=-0.7\pm0.2$ matching the observations of \citetalias{Gvozdenko2022}.
The four other star clusters found in the fiducial simulation had similarly high average [Ba/Fe] values and distribution, however, the cluster presented as the best cluster had the highest Ba abundance, in accordance with the rest of \autoref{Tab:Results}.
Some other clusters had more new star particles than the best cluster, but similar trends in the other properties.

\begin{figure*}
    \centering
    \includegraphics[width = \textwidth]{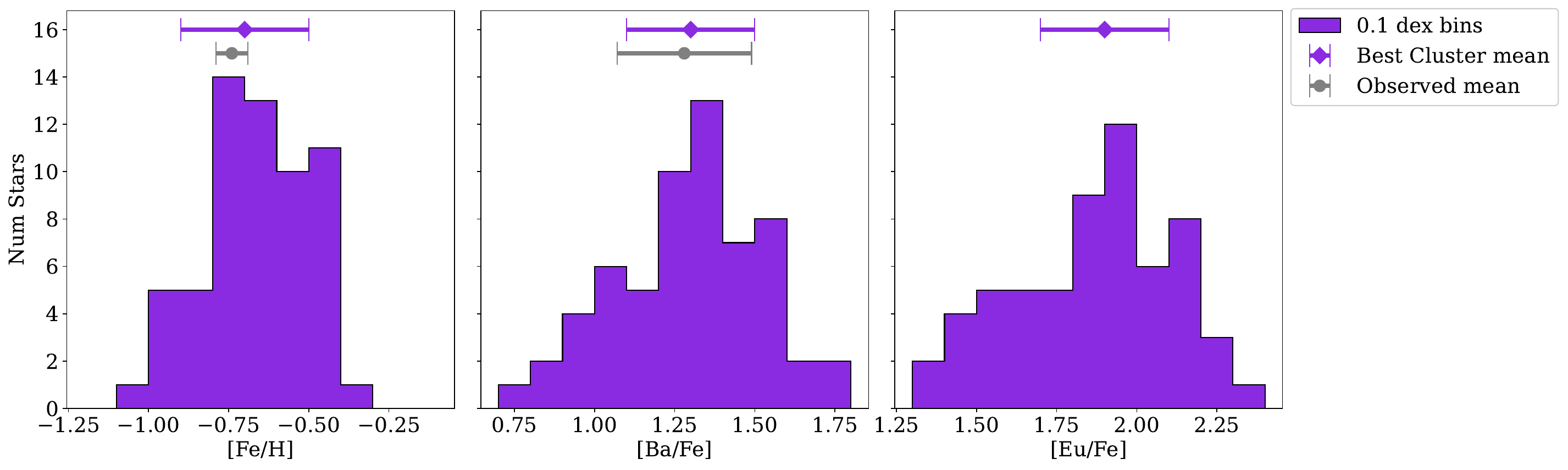}
    \caption{Histogram distributions of (left to right) Fe, Ba and Eu abundances for the best cluster are plotted here.
        A 0.1 dex bin size is used.
        The range, not the centring, of each plot is consistent at 1.1 dex for ease of comparison by eye.
        The purple diamonds with error bars are the mean value of this abundance and its standard deviation.
        The grey points are the observed values from \citetalias{Gvozdenko2022}.
        Note {[Eu/Fe]} was not measured by the observations.}
    \label{Fig:BestAbundDistr}
\end{figure*}

\autoref{Fig:BestAbundDistr} shows the abundance distributions of the best cluster for the key elements in this study.
The left panel has the [Fe/H] distribution, with the average value in agreement with the measurement in \citetalias{Gvozdenko2022} which was a requirement for selecting this as the fiducial model.
The spread of the Fe abundance is much larger than the observations, and also larger than typical [Fe/H] spreads in GCs.
The [Ba/Fe] distribution is shown in the centre panel, the average matches the observations, and the spread is of a similar size to what was found in \citetalias{Gvozdenko2022}.

The mean [Eu/Fe] of the cluster displayed in \autoref{Fig:BestAbundDistr} is 1.9$\pm$0.2.
This provides a prediction of what the Eu abundance should be in NGC1569-B, if the COL pollution hypothesis is a good model for this cluster's formation.
The large spread in Eu is not surprising, like the Fe spread is, as Ba has a large spread in the measurements that are being investigated here and both of these \textit{r}-process elements are expected to increase in tandem.
This [Eu/Fe] is higher than anything in the Milky Way for this metallicity \citep[e.g.][]{Suda2008,Tsujimoto2014}, but so was the [Ba/Fe] abundance that sparked this investigation, so a large Eu abundance was expected.
The 60 new star particles are slightly negatively skewed for both the \textit{r}-process elements, the abundance does not get much higher than the average value, but can get lower.

\begin{figure}
    \includegraphics[width=\columnwidth]{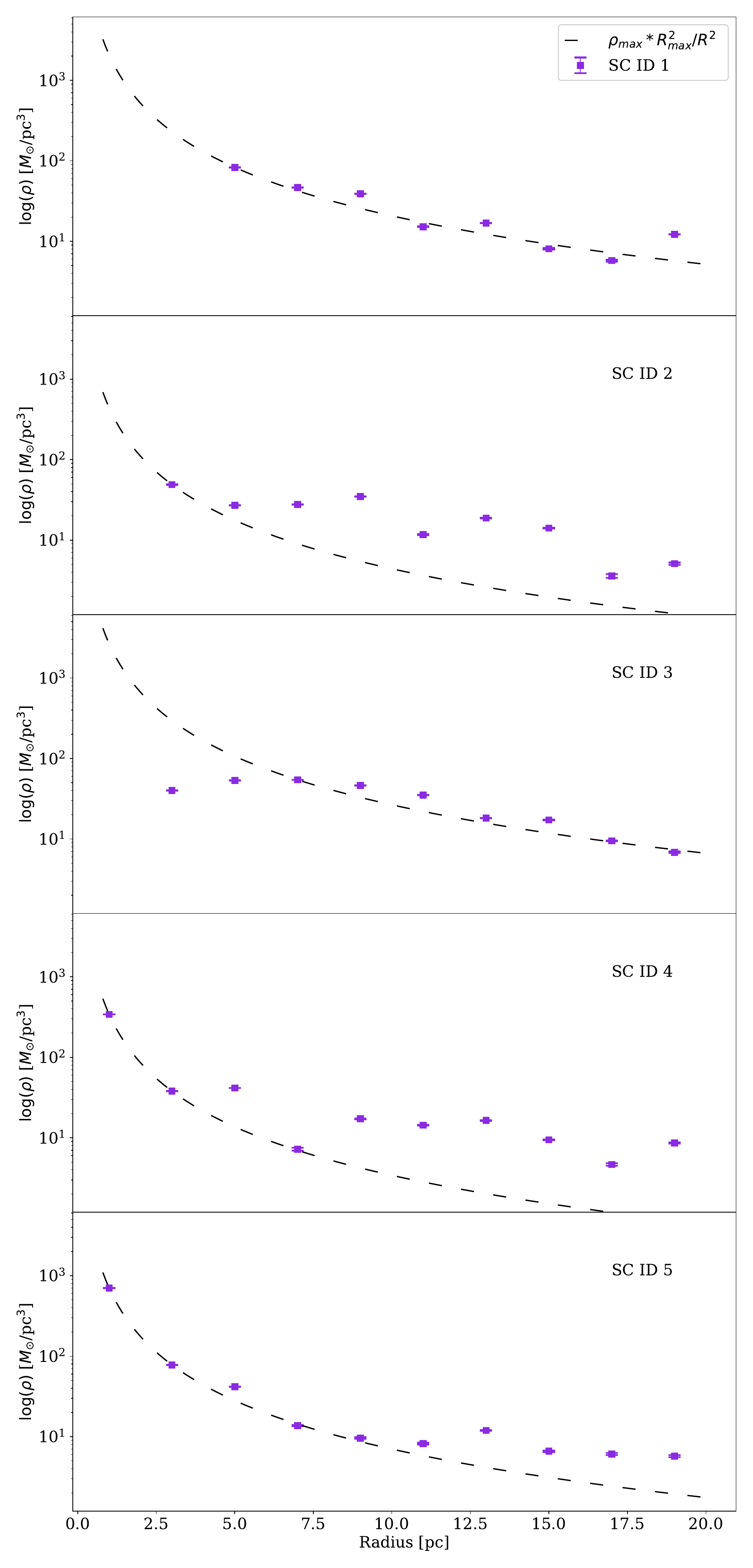}
    \caption{The density of the all the SCs in the fiducial model.
        The density is plotted on a $\log$ scale.  Each point (purple squares) are binned in \qty{2}{\pc} spherical shells.
        The error bars are $\pm\log(1+\frac1{\sqrt{n}})$, where $n$ is the number of star particles in that shell.
        The dashed black lines are the expected trend, an $R^{-2}$ relationship.
        These trend lines are scaled to pass through the point of max density, $\rho_{max}$, at  the radius of this maximum, $R_{max}$.}
    \label{Fig:AllSCDens}
\end{figure}

Only two of these five total clusters have steep enough density profiles to be real clusters.
They may have an overall negative gradient, but they do not have any central concentration of stars.

This is shown in \autoref{Fig:AllSCDens}.
The best SC presented in this work is SC ID 5, and shows a nice fit with the $R^{-2}$ trend lines.
The two lower clusters in \autoref{Fig:AllSCDens} have a non-zero density in the first shell, this includes the cluster that has the highest [Ba/Fe] of them all.
The SC ID 4 cluster has a larger-than-expected density for most of the shells, however, neither of the lower two SCs have any shells with a density less than the expected trend line.

The top three clusters in this figure have no values in their \qtyrange{0}{2}{\pc} shells, meaning there are no new star particles close to the centre of mass (COM) of the system and so cannot be real clusters.
The clusters are not as bound as their negative potential gradient implies.
They are unlikely to stay as clusters for long periods of time.

All of these clusters do not follow King profiles \citep{King1966}, their densities have no sharp edges that would be expected of older GCs.
These clusters are young, so the tidal force, from the dwarf galaxy they are embedded in these simulations, has not had time to strip the outer star particles.
There is a slight age spread in these clusters of $\sim \qty{0.5}{\giga\year}$.
This is due to a extended period of time where the SFR is high, from the first pericentre passage through to the final merging phase.

\begin{figure}
    \includegraphics[width=\columnwidth]{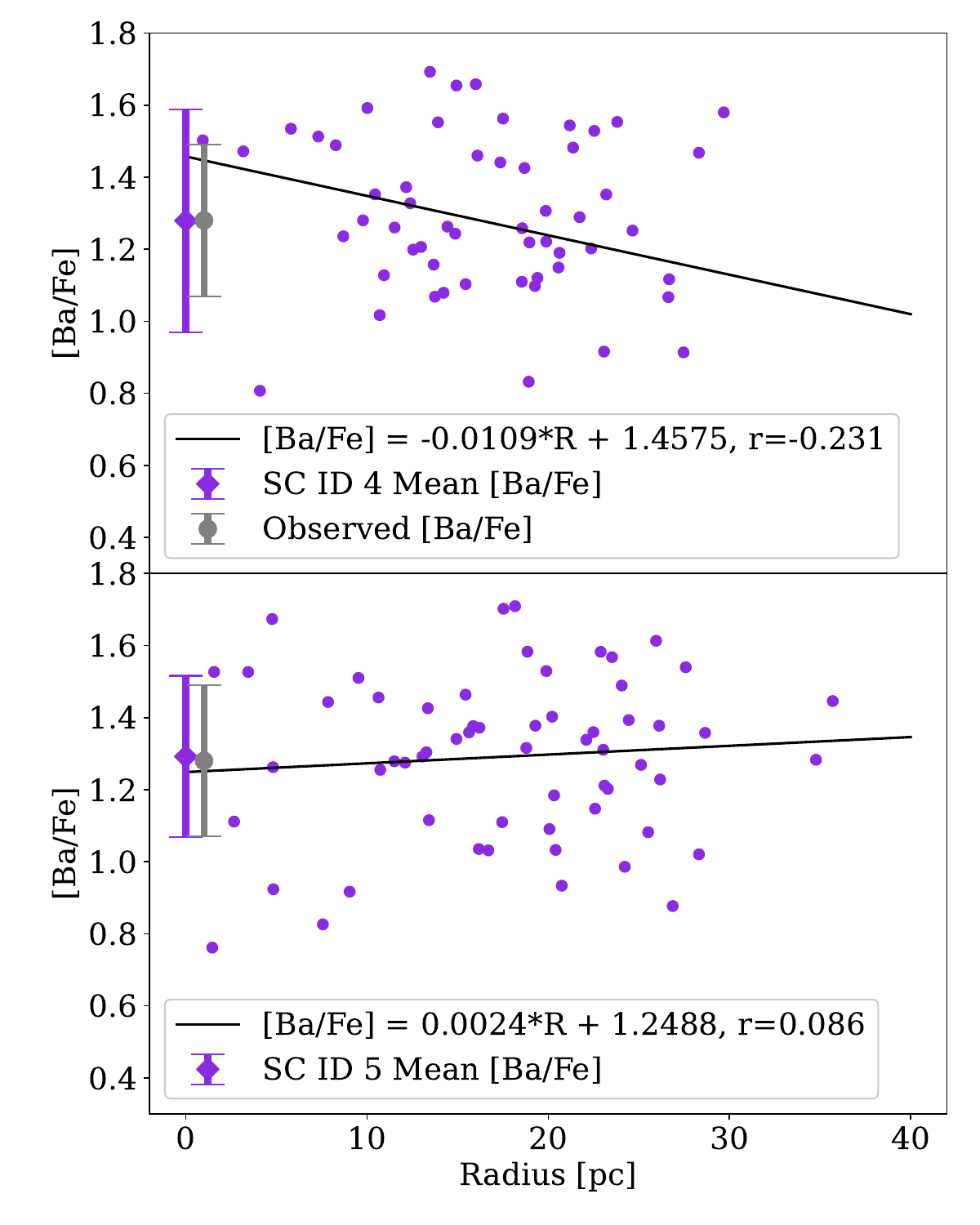}
    \caption{
        A scatter plot of the [Ba/Fe] abundance against radius from the best cluster's COM.
        The black line in the best-fitting linear regression to the points, and the legend has the equation of the line and correlation coefficient $r$.
        The purple diamond error bar is the cluster's mean [Ba/Fe] and associated standard deviation and the grey circle error bar is the reported [Ba/Fe] of NGC1569-B from \citetalias{Gvozdenko2022}, as in \autoref{Fig:BestAbundDistr}.
        The radial positions of these error bars points are just for clarity.
        The top plot is for the SC ID 4 as in \autoref{Fig:AllSCDens}, and the lower plot is for the best SC (ID 5).
    }
    \label{Fig:BaGrads}
\end{figure}

The abundance gradients of these two clusters are compared in \autoref{Fig:BaGrads}.
The best cluster has a mean [Ba/Fe] that is slightly above that of the observations, and the SC ID 4 cluster has a mean that is just lower than both of them, however, they do both agree with the measurement of NGC1569-B's Ba abundance.
SC ID 4 has a rather negative radial Ba gradient, as can be seen from the equation of the best-fitting line, and the correlation coefficient.
There is very little correlation between the radius and the Ba abundance in the best cluster, ID 5, as such there is a flat abundance gradient.
This highlights that it is the entire cluster has a high [Ba/Fe], and the average is not being raised by a few outliers.

\begin{figure*}
    \includegraphics[width=\textwidth]{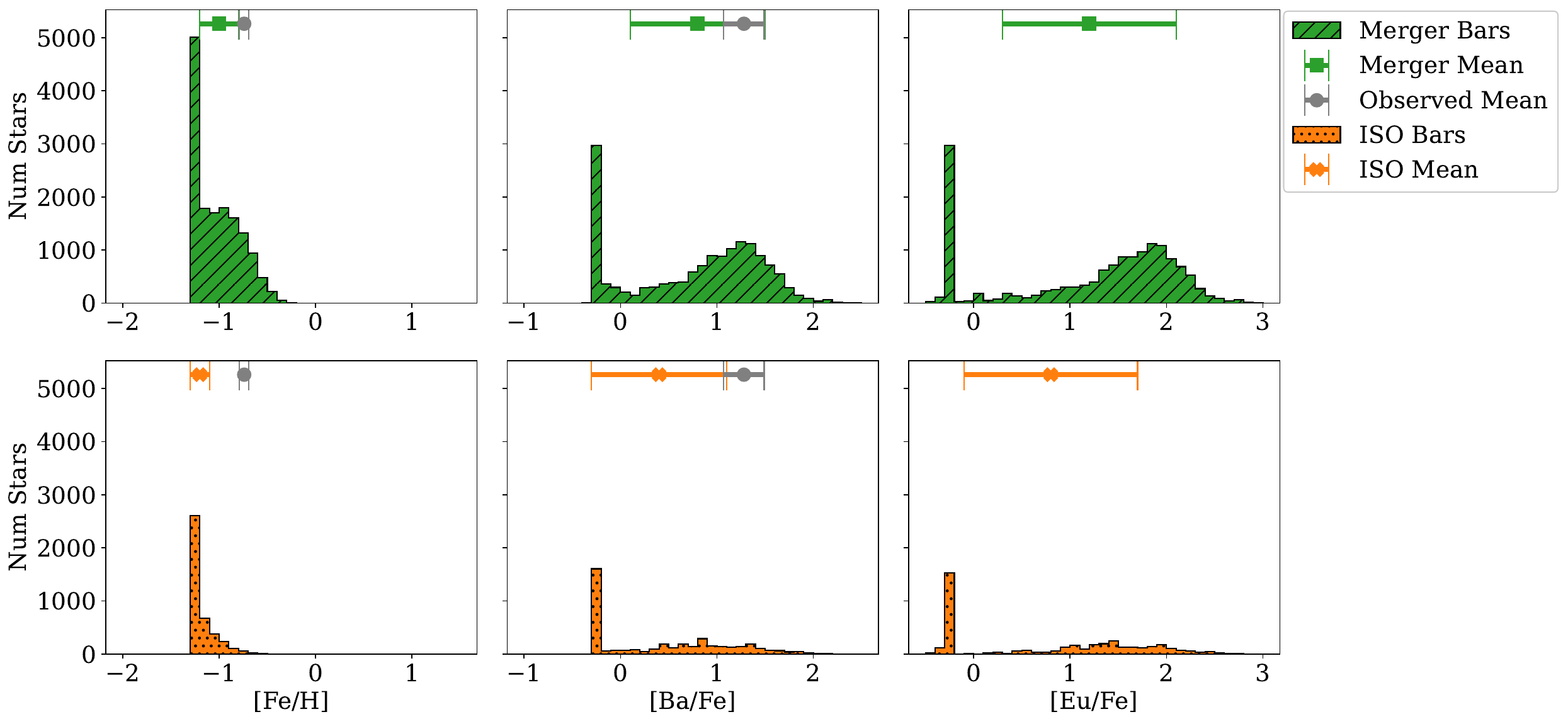}
    \caption{Similar to \autoref{Fig:BestAbundDistr} but for all the new star particles in the fiducial merger simulation M13 (hatched green, top row), not just the best cluster, and all the new star particles in the isolated galaxy simulation M16 (dotted orange, bottom row).
        The averages for the merger model is shown with the green squares, and for the isolated model in the orange  crosses.
        The bin size is still 0.1 dex.
        However, the bars appear thinner as the range of values is now 3.5 dex to show the range of [Eu/Fe].}
    \label{Fig:AllStarsBothRows}
\end{figure*}
When considering all the newly formed star particles instead of just the best cluster, similar distributions to those in \autoref{Fig:BestAbundDistr} can be made.
For the fiducial model, this is shown in the top row (hatched green) of \autoref{Fig:AllStarsBothRows}.
The iron abundance is heavily unimodal, and the number of stars at higher metallicity decreases quickly.
The mode at the minimum value is more than twice the height of the other bars.
This is contrasted by both \textit{r}-process elements,
which both show one mode at the minimum value and then peak again at around the values of the chosen cluster.
The other clusters found in this simulation, which also had rather high [Ba/Fe], are accounted for in the positive mode of these plots.
This bimodality is interesting and reinforces that the best cluster is not an anomaly, and this model of COL pollution of the \textit{r}-process is viable.
They produce a lot of these heavier elements quickly, and when so many of them occur because of a burst of SF at low metallicity there can be clusters with a high \textit{r}-process abundances when a second round of star formation happens.
The spread is reflected in the uncertainty in the mean of each element.
The [Ba/Fe] has a large spread, which encompasses the observational value, but this observed value is well-constrained enough that there is no overlap from the observational error bars, so the values do not agree.

The lowest abundances of each plot in the top row of \autoref{Fig:AllStarsBothRows} are slightly above the value set as the minimum, most clearly in the [Fe/H] plot.
The [Fe/H]\textsubscript{i} was set to -1.5 to match the observations in the cluster that formed.
The mode of the iron distribution is in the bin centred on -1.25.
There are bins with a lower metallicity than this available, but they are not populated by any new stars.
This can be explained by the evolution of the star particles that existed at the start of the simulation.
As star formation was gradual at first, as shown in \autoref{Fig:NewStarDensity}, the old star particles had time to evolve and increase the metallicity of the system before any new stars were formed.
Most of the new star particles in the peak at the minimum would have been formed in the first burst of SF, and the ones that form the clusters are in the higher peaks of the \textit{r}-process elements.

\subsection{Parameter Dependence}\label{SubSec:InputSens}

Having found a fiducial model (M13 of \autoref{Tab:Results}) that matched the observations of \citetalias{Gvozdenko2022}, our attention turned to understanding the effects of the assumptions of the other parameters of the simulations.
These tests show that the parameters that had been assumed have at most a small effect on the results, shown in \autoref{Tab:Results} for models M14--M22.
These models have the same CCSN per $\unit{\Msun}$, and CCSN per COL as the fiducial model, the changes made to each model are detailed in the ``Other Changes'' column of \autoref{Tab:Results}. We now describe each of the changes, and their effects, in turn.

The largest effect is decreasing the threshold gas density, $\rho_{th}$, to 100 cm\textsuperscript{-3}.
As shown by M14 in \autoref{Tab:Results}, this changes the [Fe/H] significantly, [Fe/H]\textsubscript{f} $=-0.5 \pm 0.2$, placing on the upper edge of the uncertainty limit from the fiducial model.
This cluster did require a slightly larger NR to be detected, \qty{52.5}{\pc} which is 1.5 times larger than the standard \qty{35}{\pc}. M14 has more than twice as many new star particles form than the fiducial simulation, yet they aren't clustered.
Only one cluster was formed at this larger NR, it passes the potential gradient check but does not have any star particles within \qty{10}{\pc} of the COM, making the shell density against radius rather empty.
It seems the lower $\rho_{th}$ makes it easier for stars to form, as expected, but harder for clusters to form.
This may be due to the density of nearby gas particles influencing each other.
It is more likely that a group of particles will all approach the higher density together, encouraging cluster formation.
While perhaps an individual particle can reach the lower value on its own and the feedback from the new young stars could lower the density of the surrounding gas, which decreases clustering tendencies.

The initial mass ratio of the two galaxies was also investigated.
For M15, the ratio was changed to 1:0.5.
This does lower the total mass in the system, by a factor of $\frac14$ for the ratio used.
However, it seems to not affect the [Ba/Fe] value, indicating that the SF is brought on by the merger and the total mass of the system is less important.
Three clusters pass the negative potential gradient check, but none have a non-zero density near the COM.

The `ISO' tag in M9 and M16 means there was no merger initialised, just leaving the one dwarf galaxy to evolve on its own.
These models failed to produce a cluster, even at an extreme NR of \qty{157.5}{\pc}, more than 4 times the standard value.
With nothing to promote SF, it makes sense that there would not be a new cluster of stars in this small galaxy. The bottom row (dotted orange) of \autoref{Fig:AllStarsBothRows} shows the abundance distribution  of the new stars in the isolated model with otherwise fiducial parameters, M16.
There are 3 times fewer new star particles than in the merger model (top pannel), the majority of these stars are at the minimum abundance for each element.
This demonstrates that a merger drives the starburst and leads to cluster formation as no cluster was found in M16 even with a large NR.

The pericentre distance is standard at \qty{8.75}{\kilo\pc}, models that vary just this parameter (M17, M18, M21 and M22), show good agreement with the fiducial model.
This means the result of the fiducial simulation is rather invariant under a change of pericentre of the merger.
The changes in both the barium and iron abundances are only 0.1 dex, which is within the uncertainties of these measurements.
There seems to be no trend in the number of SCs that pass the negative potential gradient check and the pericentre distance.
Only in M21 is there no SC that has a well-fitting $\rho \propto R^{-2}$ cluster, the rest of the models had at least one.

The pericentre changes can be combined with a mass ratio change.
The `--' of M19, from no grouping of new star particles passing the negative potential gradient check, is interesting because it combines two changes that both successfully formed clusters.
The larger pericentre of M20 with the lower mass ratio had outputs in agreement with the fiducial model, so there should not be too large of an effect for any changes in these parameters on the abundances.
However, this model only had one cluster pass the potential gradient check, and it has no star particles within the first \qty{2}{pc}.
It seems that the stars will form with high Ba abundances, but the merger parameters seem to disfavour clustering.

The only 0.1 dex variance in the abundance outputs as sensitivity to any of the merger parameters from these models is a positive result.
The lack of clustering in some of these models is of slight concern.
However, this work did not attempt to find any exact orbital merger history of NGC1569.
Instead, the merger was used to excite the high SFR,  needed for COL pollution, as seen in the bursty SFH of NGC1569 \citep{Angeretti2005}.
The invariance shown in the abundances solidifies that it is the COL rates that are important, not the specific merger parameters.
The mergers simulated here have old star particles in both bodies, but there is no \textit{a priori} reason that this is required.
This merger need not be with a galaxy, it could just be an HI cloud, which fits slightly better with the results of \citet{Johnson2013}.

The M23 was designed to show the degeneracy between the CCSN per $\unit{\Msun}$ and CCSN per COL values.
The COL per $\unit{\Msun}$ value, the lurking variable in these simulations as it is not an explicit input, are the same, $1\times10^{-4}$.
Interestingly, the value of this lurking variable is such a round number.
The flag of this model in \autoref{Tab:Results} indicates congruence with M13.
I.e. the same number of new star particles and clusters were formed in both.
It is not just that the [Ba/Fe] is the same, they were the exact same simulation, despite the difference in input parameters.
This was the expected behaviour because this run was designed to have the same value of COL per $\unit{\Msun}$ for a different IMF, one that is physically motivated, and derived from \autoref{Eq:AlpSSFR}.
The fiducial model does have the benefit that it has a standard \citep{Salpeter1955} IMF.
It shows that the fiducial simulation is not the only set of parameters that works.
Any combination of CCSN per $\unit{\Msun}$ and CCSN per COL that have a ratio of $1\times10^{-4}$ would give the same result.
It is interesting to note that this model has a CCSN per COL that matches the current GRB rate estimates, as described in \autoref{SubSec:ColParams}.

\section{Discussion}\label{Sec:Dis}

\subsection{Unresolved problems in COL scenario}\label{SubSec:ColProbs}
Although we have provided a reasonable explanation for the observed high [Ba/Fe] of NGC1569 B and theoretical predictions (e.g., [Eu/Fe]) of other properties of the SC,  we here list several unresolved problems of the proposed COL scenario.

\subsubsection{Very high [Cr/Fe] and [Sc/Fe]}

There are other anomalous abundances of NGC1569-B as reported in \citetalias{Gvozdenko2022}.
The [Cr/Fe] $=0.50\pm0.11$ and the [Sc/Fe]$=0.78\pm 0.20$ are both higher than expected, but not nearly as extreme as the [Ba/Fe] measurement that our study has focused on.
The high [Cr/Fe] could be explained if the yield of Cr in COLs are close to that of MR-SNe yields for this element.
\citetalias{Gvozdenko2022} does state that the Sc value is particularly uncertain.
The spread is discussed at length with the changes to the measurement when fitting fewer Sc lines shown, but if the measurement is to be believed then the abundance contributions from massive stars, type Ia, and type II supernova would have to be different from the Milky Way and other galaxies.
The spread in [Ba/Fe] is not mentioned in the text merely shown in a table in the appendix.
COLs have a difficult job explaining this Sc over-abundance analytically.

\citet{Iliadis2016} show that super-AGB stars or novae of CO or ONe white dwarfs are sites that can produce high Sc pollution.
However, the timescale of the white dwarf novae is slightly too long for NGC1569-B; at least \qty{100}{\mega\year}.
This is longer than the current age of the cluster, and having earlier SF do this pollution would be challenging because of the spatial extents of these bursts.
We suggest that these super-AGB stars can pollute the ISM before cluster formation.
Much like the COLs, these stars would have been formed in the earlier round of starburst, super-AGB stars are active on timescales \qty{30}{\mega\year}, slightly longer than COLs which casts doubt on whether they would be a good candidate.
Super-AGB stars would also force the K abundance to be high, as the Mg-K anticorrelation presented in \citet{Iliadis2016} would take the [Mg/Fe] $0.22\pm0.05$ of \citetalias{Gvozdenko2022} to a [K/Fe] of about 1.5.
\citetalias{Gvozdenko2022} did not measure the [K/Fe] of NGC1569-B, so there is currently no way of knowing this abundance.
A mixture of COL and super-AGB ejecta could be an interesting explanation of the mix of high values of Sc, Cr and Ba abundances.

\subsubsection{Fe spread}

The final [Fe/H] value for the fiducial model is $-0.7\pm0.2$, and this uncertainty of 0.2 dex is a large dispersion for a Fe abundance among the new star particles in the cluster.
The left panel of \autoref{Fig:BestAbundDistr} shows this larger scatter in the [Fe/H] values in histogram form.
\citetalias{Gvozdenko2022} measured [Fe/H]$=-0.74\pm0.05$, with the $\pm0.05$ being a measurement uncertainty, not a dispersion, as the integrated light spectra technique used does not measure the [Fe/H] (or any abundance) for individual stars.
The COL model seems to produce this large spread in Fe naturally.
The AGB pollution discussed above would give a small Fe spread, as these stars have no Fe in their ejecta.
If the timescales worked in their favour, (super-)AGB stars could be preferred over COLs, but NGC1569-B is too young.
The high spread in [Fe/H] is a prediction of the model. While observations of this dispersion would be challenging due to technological limitations, any measurements made would be useful to check the validity of the model.

\subsubsection{Parameter degeneracy}
The number of COLs that occur is the important value, but it is highly degenerate.
Two ways of increasing this number have been explored, changing the CCSN per COL input and changing the IMF, achieved by changing the CCSN per $\unit{\Msun}$ input.
The former increases this total number of COLs because COLs are more common, observationally this is linked to the number of long GRBs.
The latter is increasing the number of CCSNe, which can result in the same number of COLs with a lower fraction of CCSN being COLs.
This is dependent on the IMF power law exponent changing with SFR, which there is some evidence for \citep{Gunawardhana2011}.

The degeneracy between these parameters is very strong, as discussed above the lurking variable of COL per $\unit{\Msun}$ explains the degeneracy.
There is hope to break it, but not through these simulations, only with further study into both SFR-dependent IMFs and GRB frequencies.
If it is found that a low CCSN per COL is acceptable, meaning there are more long GRBs observed, then a \citet{Salpeter1955} IMF works well with these results, as the observations can be reproduced with a CCSN per COL of 70, the value of this parameter in the fiducial model, about $\frac14$ the estimated rate (see \autoref{SubSec:ColParams}).
The higher GRB rate may only need to be acceptable at lower metallicities.
The current estimates are for the local universe, which is not metallicity specific.
If there are the current estimates on the GRB frequencies are overestimates, i.e. they predict more GRBs than are observed even at low metallicities, an SFR-dependent IMF is needed to reproduce the observed [Ba/Fe], as seen in M23.

\subsubsection{Origin of High COL fraction}
It remains unclear why rotating massive stars can be formed efficiently in a dwarf galaxy merger environment.
The CCSN per COL value of the fiducial model (70) is towards the lower end of the physically applicable range.
This is about a factor of 4 lower than the current GRB estimates would have, which is $\sim270$ (see \autoref{SubSec:ColParams}).
One has to wonder how these COLs would have formed so efficiently.
Lowering the initial metallicity of the system does improve the chances of massive stars being COLs due to previously discussed reasons to do with the predicted rotation rates of WR stars (see \autoref{SubSubSec:Cols}).
At such an initial [Fe/H] to reproduce the observed [Fe/H] of NGC1569, the CCSN per COL values used in the fiducial simulation of 70 are well within the range predicted by \citet{Heger2003} of $\sim \qty{10}{\percent}$ of CCSN being COLs.
The high SFR of the simulations at a low metallicity also helps the model as all the COLs can form in optimal conditions, and the star clusters form afterwards.
The SFH of NGC1569-B as described by \citet{Angeretti2005}, with three intense bursts in recent history, is consistent with these results.
The first burst would have formed the COLs, and the latest would have formed the cluster.

\subsection{Is NGC1569-B unique?}\label{SubSec:UniqueSSC}

Milky Way GCs \citep[e.g.][]{Larsen2017} and those in dwarf galaxies \citep[e.g.][]{Larsen2012,Larsen2014,Gvozdenko2024} have [Ba/Fe] much lower than observed in \citetalias{Gvozdenko2022}, when all were observed with the same technique of an integrated light spectrum.
This suggests that the COL pollution proposed here is unique to NGC1569-B.
The GCs in our galaxy could have been formed in isolation during dwarf galaxy formation before being captured into the Milky Way, and this difference in the SF environment could go some way to explaining the difference.
The high CCSN per COL fraction needed for the fiducial model presented in this work is questionable to have only occurred once, but the same problem exists for any explanation of such a seemingly unique abundance.
Our presented CCSN per COL value, 70, being only a factor of about 4 lower than the GRB estimated rate puts it well within the bounds of possibility.
Observations of long GRB host galaxies could be crucial, if the hosts preferentially show evidence of mergers then the COL fraction could be higher after a merger.

The bursty SFH of NGC1569 may be such that there was the correct number of COLs formed in the lifetime of the GMC.
This can be seen in greater significance by using the parameter of the fiducial model, $f_{\text{COL}}=70$,  in \autoref{Eq:CCSNfromCOL}, \autoref{Eq:MassCol} and \autoref{Eq:TimeFromSFR}.
The values become $M_{*\text{, COL}}=\qty[separate-uncertainty=true]{1.8(1.2:0.7)e6}{\Msun}$ and $\Delta t_{\text{SF, COL}}=\qty[separate-uncertainty=true]{3.8(2.5:1.6)}{\mega\year}$.
These are even more in favour of the COL model than our original estimates, with the SF time being well less than the lifetime of a GMC.
This is without even considering the complicating effects of lowering the initial metallicity, which lowers the required number of COLs and makes them more likely.
Therefore, SF before the -B cluster formed is certainly enough to pollute the GMC.
The unique circumstances of the merger into a bursty SFH and then forming an SSC that is bright enough to observe from Earth are all compounding factors into this [Ba/Fe] being so high. The other GCs mentioned above may have formed with less erratic histories and thus a lower Ba abundance as there would have not been enough SF right before their GMCs collapsed to have enough COLs occur for the \textit{r}-process pollution.

\subsection{Future work}\label{SubSec:Future}

There was no attempt to make mock observations of the clusters detected in the simulations, these would be a good progression to this work to see if the results are still valid.
By taking an integrated light spectrum of the clusters identified in the simulations, a more self-consistent comparison of the abundances can be made.
Any systematic errors would be present in both the real and the mock spectra, and the biases of taking the mean of all the star particles could be removed.
While averaging the abundance of each star particle in principle has a similar effect to a spectrum of all the light, no weighting of the intrinsic brightness of each star is taken into account.
The colour-magnitude diagram (CMD) binning that the integrated light spectra analysis performs does take stellar spectral type into account \citep{Larsen2012}.
This would be the difficult step of taking these mock measurements, as the star particles in the simulation represent more than one star.
The CMD binning makes less sense at this resolution, and finding a way around this problem is left for future work as it is outside the scope of this work.

Taking the average is not the worst choice we could have made, instead the highest single star particle's [Ba/Fe] could have been chosen for each cluster.
As shown in \autoref{Fig:BestAbundDistr} that would lead to $\sim 0.5$ dex higher abundances, which would have skewed the input parameters to prefer fewer COLs.
While we are confident in our results from the averages alone, the mock observation technique would be a good extension.

Another opportunity for more research would be taking follow-up observations of NGC1569-B and -A.
This work predicts the -B cluster to have a [Eu/Fe]$=1.9\pm0.2$ and an attempt to verify this prediction is needed to determine if these simulations accurately describe the cluster.
The COL pollution model as a concept would predict the -A cluster would also have a large \textit{r}-process abundances, as the initial conditions of the ISM and the merger and SFH are the same.
The simulations back this theory up, with most models having a few more clusters with [Ba/Fe] within 0.2 dex of the presented cluster.
Performing these observations is not in our expertise, but any attempts to do so are welcome.

\section{Conclusion}\label{Sec:Conc}
%Intro Para
The massive star cluster NGC 1569-B is observed to have unique chemical abundance patterns in its stars, in particular, extremely high [Ba/Fe] of $1.28\pm0.21$ \citepalias{Gvozdenko2022}.
We consider that this Ba is of \textit{r}-process origin due to chemical enrichment by COLs, NSMs or MR-SNe.
Accordingly, we have investigated the required  number of each of these \textit{r}-process sites and some properties of the SF needed to produce them.
We then investigated the star cluster formation from merging dwarf galaxies in the context of the COL scenario using supercomputer simulations of the chemodynamical evolution, with COLs being the \textit{r}-processes synthesizers.

\begin{enumerate}
    \item The number of events needed to pollute a GMC up to the correct [Ba/Fe] levels vary between the different \textit{r}-process sites.
          Collapsars requires $M_{*\text{,COL}}=\qty[separate-uncertainty=true]{2.6(1.7:1.1)e7}{\Msun}$ of stars need to form, and this takes $\Delta t_{\text{SF, COL}}=\qty[separate-uncertainty=true]{53(36:22)}{\mega\year}$ at the current SFR \citep[\qty{0.48}{\Msun\per\year}][]{Hunter2004}. This time is consistent with the SFH of NGC1569 \citep{Angeretti2005}, which implies COLs are a good candidate for the \textit{r}-process pollution of NGC1569-B.\\

    \item The number of MR-SNe needed is high, and so the corresponding mass of stars that need to form and the time of SF is inconsistent with the SFH of the galaxy.
          The old stellar population in NGC1569 is expected to have 84 NSMs occur during the lifetime of the GMC that formed the -B cluster, when the number required for the Ba pollution is orders of magnitudes greater.
          Therefore, we reject these other sites for the \textit{r}-process pollution of NGC1569-B and conclude that COLs must do this pollution. \\

    \item Star clusters can be formed in dwarf galaxy mergers, and we had a cluster match the observed properties of NGC1569-B.
          The main result was that a [Ba/Fe]$=1.3\pm0.2$ and a [Fe/H]$=-0.7\pm0.2$, matching the \citetalias{Gvozdenko2022} observation.
          This was achieved with a standard \citet{Salpeter1955} IMF, an initial metallicity of [Fe/H]\textsubscript{i}$=-1.5$ and a CCSN per COL of 70, which is only a factor of 4 lower than current long GRB rates would put it.
          The best cluster has a negative potential gradient, as was required, and a slight negative skew in the \textit{r}-process element distributions.
          We have shown these findings are robust to changes in the parameters of the dwarf galaxy merger, but that a merger of some kind is necessary. \\

    \item NGC1569-B did not have an Eu abundance measurement made. However, our simulations did measure the [Eu/Fe] of the clusters.
          From the fiducial model, a prediction of the Eu abundance of the real SSC is made; [Eu/Fe] = $1.9\pm0.2$.
          Follow-up measurements of Eu are needed to confirm or refute this result.\\

    \item The degeneracy between the simulation input parameters of CCSN per \si\Msun and CCSN per COL was the main issue with this study.
          Another cluster fits the observations with an SFR derived IMF and an observational CCSN per COL value.
          The lurking variable of COLs per \si\Msun is what controlled our results, even though it was not an explicit input, but there is hope to break the degeneracy with more work.  \\

    \item While we are confident in our results, there is still work to be done.
          A mock observation of the simulations to get a better comparison to real world data through integrated light spectra would be a great step to take.
          To achieve this a consideration of the CMD of a simulated cluster is required.\\
\end{enumerate}

\section*{Acknowledgements} \label{Sec:Acknow}
The authors would like to thank the anonomous reviewer for their helpful coments.
BL would like to thank ICRAR for the opportunity to be a part of their summer studentship program.
T. T. acknowledges the support by JSPS KAKENHI Grant No. 23H00132.

%OzSTAR asks for:
This work was performed on the OzSTAR national facility at Swinburne University of Technology.
The OzSTAR program receives funding in part from the Astronomy National Collaborative Research Infrastructure Strategy (NCRIS) allocation provided by the Australian Government, and from the Victorian Higher Education State Investment Fund (VHESIF) provided by the Victorian Government.

%%%%%%%%%%%%%%%%%%%%%%%%%%%%%%%%%%%%%%%%%%%%%%%%%%
\section*{Data Availability} \label{Sec:DataAvail}
The data used for this work will be made available on reasonable request.

\bibliographystyle{mnras}
\bibliography{bibfile}

%%%%%%%%%%%%%%%%% APPENDICES %%%%%%%%%%%%%%%%%%%%%
%\appendix
%%%%%%%%%%%%%%%%%%%%%%%%%%%%%%%%%%%%%%%%%%%%%%%%%%

% Don't change these lines
\bsp	% typesetting comment
\label{lastpage}
\end{document}